\documentclass[12pt]{iopart}
\usepackage{graphicx, color}
\usepackage{color}
\usepackage{natbib}
\usepackage{indentfirst}
\usepackage{url}
\usepackage{amssymb}
\eqnobysec

\makeatletter
\def\subsubsection{\@startsection {subsubsection}{3}{\z@}{2.5ex plus -1ex minus-.2ex}
{1ex plus 1.2ex}
{\small\it}
}

\begin{document}

\title[An Origin for the Angular Momentum of Molecular Cloud Cores]{An Origin for the Angular Momentum of Molecular Cloud Cores: a Prediction from Filament Fragmentation}

\author{\small Yoshiaki Misugi$^1$, Shu-ichiro Inutsuka$^1$ and Doris Arzoumanian$^{1,2}$}
\address{$^1$ Department of Physics, Graduate School of Science, Nagoya University, 464-8602 Nagoya, Japan}
\address{$^2$ Centro de Astrofisica da Universidade do Porto, Rua das estrelas, 4150-762 Porto, Portugal}

\begin{abstract}
The angular momentum of a molecular cloud core plays a key role in star formation, since it is directly related to the outflow and the jet emanating from the new-born star and it eventually results in the formation of the protoplanetary disk. However, the origin of the core rotation and its time evolution are not well understood. Recent observations reveal that molecular clouds exhibit a ubiquity of filamentary structures and that star forming cores are associated with the densest filaments. Since these results suggest that dense cores form primarily in filaments, the mechanism of core formation from filament fragmentation should explain the distribution of the angular momentum of these cores. In this paper we analyze the relation between velocity fluctuations along the filament close to equilibrium and the angular momentum of the cores formed along its crest. We first find that an isotropic velocity fluctuation that follows the three-dimensional Kolmogorov spectrum does not reproduce the observed angular momentum of molecular cloud cores. We then identify the need for a large power at small scales and study the effect of three power spectrum models. We show that the one-dimensional Kolmogorov power spectrum with a slope $-5/3$ and an anisotropic model with reasonable parameters are compatible with the observations. Our results stress the importance of more detailed and systematic observations of both the velocity structure along filaments and the angular momentum distribution of molecular cloud cores to determine the validity of the mechanism of core formation from filamentary molecular clouds.
\end{abstract}


\section{Introduction}
\begin{figure}[t]
\centering
\includegraphics[width=8cm]{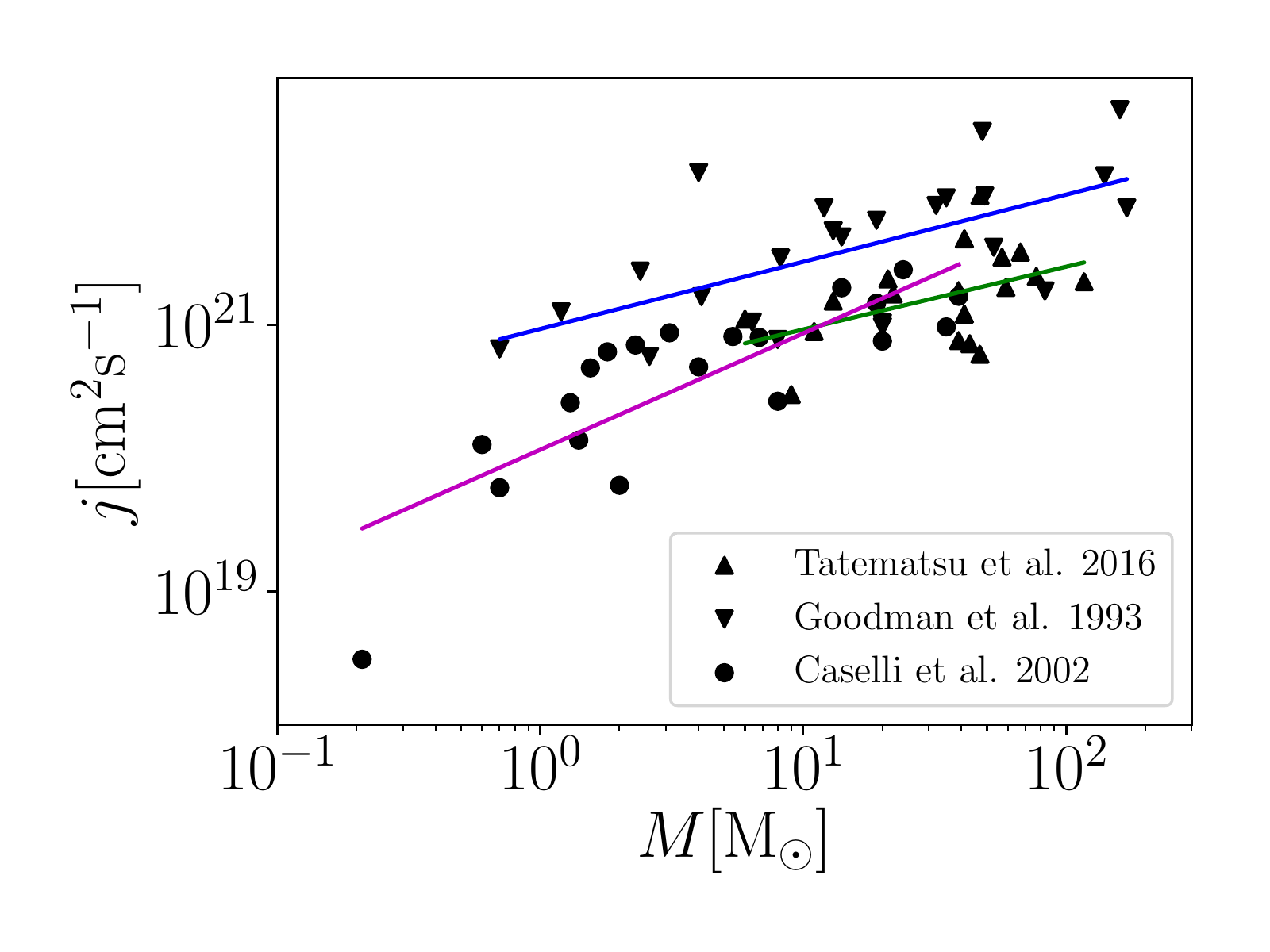}
\caption{Observations of angular momentum of molecular cloud cores. The horizontal axis is the core mass and the vertical axis is the core specific angular momentum (see text for details). The up-triangles, down-triangles, and circles are observational data from \citet{Tatematsu2016}, \citet{Goodman1993}, and \citet{Caselli2002}, respectively. The green, blue, and magenta solid lines are the results of the fits by applying the least square method to the data of \citet{Tatematsu2016}, \citet{Goodman1993}, and \citet{Caselli2002}, respectively. The values of the slopes of the fits to the data of \citet{Tatematsu2016}, \citet{Goodman1993}, and \citet{Caselli2002} are 0.5, 0.5, and 0.9, respectively.}
\label{goodmancasellitatematsu}
\end{figure}

The angular momentum of molecular cloud cores plays an essential role in the star formation process, since it is at the origin of the outflow and the jet, results in the formation of the protoplanetary disk, and defines the multiplicity of a stellar system (single star, binary, multiple stars). The angular momentum of a core is defined at the initial conditions of the core formation \cite[e.g.,][]{Machida2008} and understanding how molecular cloud cores obtain their angular momentum is a key question in the star and planet formation studies. The angular momentum of cores has been derived from molecular line observations using the ${\rm NH_3}$ transition \citep{Goodman1993} and ${\rm N_2H^+}$ line \citep{Caselli2002}. More recently, \citet{Tatematsu2016} studied a sample of cores in the Orion A cloud and derived their rotation velocity using ${\rm N_2H^+} J=1-0$. These observational results have mainly shown two properties of the specific angular momentum of cores $j$ (angular momentum per unit of mass). 1) The range of the specific angular momentum of the cores is $j = 10^{20 - 22} {\rm cm^2 s^{-1}}$, and 2) it is a weakly increasing function of the core mass $M$, $j \propto M^{0.5-0.9}$. These results are shown in Figure \ref{goodmancasellitatematsu} (cf., Section 2.1). \par 
Recent results derived from the analysis of Herschel data revealed that stars mainly form in filamentary structures \citep{Andre2010,Arzoumanian_2011,Konyves2015} and the characteristic width of filaments is $\sim 0.1$ pc \citep{Arzoumanian_2011,Arzoumanian2019,Koch2015}. Moreover, observations show that prestellar cores and protostars form primarily in the thermally critical and supercritical filaments (${\rm M_{\rm line}} \gtrsim {\rm M_{\rm line,crit}}$) \citep{Andre2010,Tafalla_2015}. The line mass ${\rm M_{\rm line}}$ of filaments is important for the star formation process. Theoretically, a thermally supercritical filament is expected to be the site of self-gravitational fragmentation and the birth place of star forming cores \citep{Inutsuka_Miyama1997}. Therefore, if most of the star forming cores are formed along critical/supercritical filaments, the theory of core formation out of filament fragmentation is expected to explain the origin of angular momentum of the cores. \par
The three-dimensional velocity structure along the filaments is needed to infer the distribution of the angular momentum of cores. However, the line of sight component of the velocity is solely accessible from molecular line observations. Note that the line of sight velocity may be dominated by the velocity perpendicular to the axis of filament when the line of sight is nearly perpendicular to the filament axis. The fluctuations of the centroid velocity of filaments close to equilibrium (${\rm M_{\rm line}} \sim {\rm M_{\rm line,crit}}$) is observed to be sub (tran) sonic \citep{Hacar2011,Arzoumanian2013,Hacar2016} . While deriving the velocity power spectrum from molecular line data is observationally difficult, the power spectrum of the column density fluctuation has been already measured in the interstellar medium (ISM) \citep{Miville2010,Roy2019} and along a sample of filaments observed by Herschel \citep{Roy2015}. These observations reveal that the power spectrum of the column density fluctuation is consistent with the power spectrum generated by subsonic Kolmogorov turbulence down to the scales of $\sim 0.02$ pc in the nearest regions (distance $\sim 140$ pc).\par
In this paper, we try to establish the relation between velocity fluctuations along the filament and the angular momentum of the cores that result from the fragmentation of the parent filament. Using the available observational data we examine whether we can explain the origin of the angular momentum of cores as a result of the filament fragmentation process.\par
The structure of the paper is as follows: The method of our calculation is given in Section 2, in Section 3 we show our results. The comparison of the derived models is given in Section 4. Section 5 presents a discussion and suggests implications of our results in the context of star formation. We summarize this paper in Section 6.

\section{Analyses}
In this section, we first mention the observational data used in this paper. Then, we explain the method for calculating the angular momentum of molecular cloud cores formed by filament fragmentation.
 

\subsection{Core Angular Momentum Derived from the Observations}
Observationally, the specific angular momentum, $j=pRv_{\rm rot}$ (angular momentum per unit of mass) of a molecular cloud core may be derived from the observed radius $R$ of the core, the rotational velocity $v_{\rm rot}$, and $p$, which is related to the density profile of the core. For a uniform density sphere, $p=2/5$, while for a singular isothermal sphere $p=2/9$. Figure \ref{goodmancasellitatematsu} shows the relation between the specific angular momentum and their mass for a sample of cores studied by \citet{Goodman1993}, \citet{Caselli2002}, and \citet{Tatematsu2016}. Since the specific angular momentum of the cores is not given in \citet{Caselli2002}, we derived them using $j=pRv_{\rm rot}$, with the values of $v_{\rm rot}$ from Table 5 in \citet{Caselli2002}, $R$ from their Table 3, and $p=2/5$, which is the same value used in \citet{Goodman1993}. For the core masses, we used the $M_{\rm ex}$ values given in Table 4 of \citet{Caselli2002}. Note that Figure \ref{goodmancasellitatematsu} includes very large objects with sizes up to 0.6 pc, which may be clumps instead of real cores \citep{Goodman1993}. While the sample of \citet{Goodman1993} includes very large objects, \citet{Caselli2002} and \citet{Tatematsu2016} do not include such clumps. Our conclusion do not depend on whether these large objects are included or not. 

\subsection{Filament Setup}
\begin{figure}[t]
\centering
\includegraphics[width=8cm]{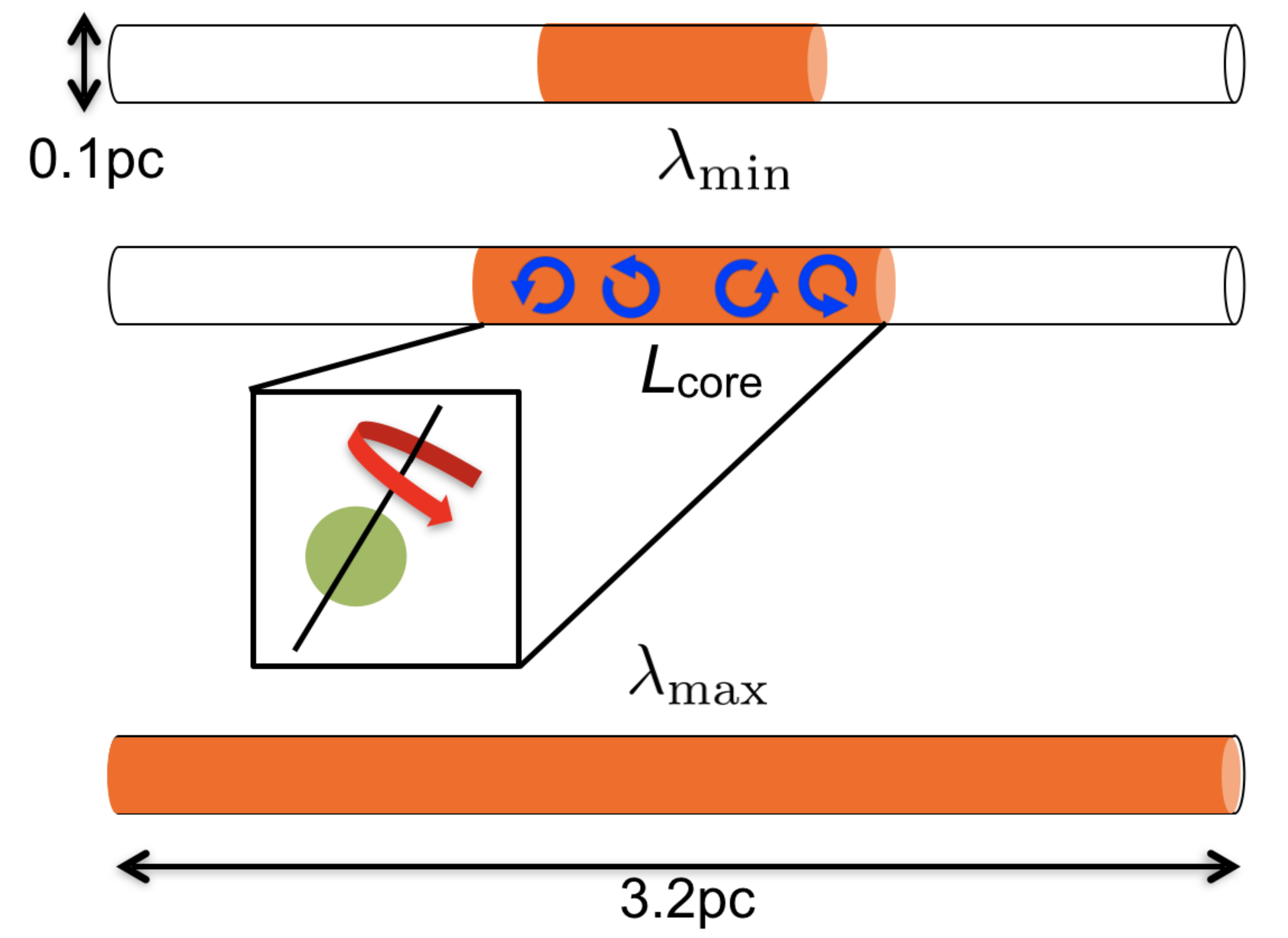}
\caption{Schematic figure describing the calculation of the angular momentum of a core. We calculate the angular momentum and the mass within model cores defined by their length (see Section 2.4). We vary the position of the cores and their length between $\lambda_{\rm min}$ and $\lambda_{\rm max}$ to generate the model $j-M$ relation.}
\label{Alfig1}
\end{figure}

We consider an unmagnetized isothermal model of filament with a width of 0.1 pc \citep{Arzoumanian_2011,Arzoumanian2019}, a uniform density, and a line mass equal to the critical line mass, ${\rm M}_{\rm line,crit}=16 \ {\rm M_{\odot} pc^{-1}}$ for $T=10$ K. An isothermal gas filament with line-mass larger than the above value cannot be in equilibrium, without non-thermal support such as magnetic field or internal turbulence.\par
We define the $z$-axis along the main axis of the filament and the $x$- and $y$-axis the transverse directions (see Figure \ref{Alfig1}). We use 32 grid cells in $x$ and $y$-axis, and 1024 grid cells in $z$-axis. We use the periodic boundary condition in this domain.
The length of the filament is $L_z=3.2$ pc, and the total mass of the filament is ${\rm M}_{\rm max}={\rm M}_{\rm line,crit} L_z$. $M={\rm M}_{\rm line,crit} L_{\rm core}$ is a mass of a section of the filament where $L_{\rm core}$ is an arbitrary length $\lambda_{\rm min}< L_{\rm core} <\lambda_{\rm max}$, where $\lambda_{\rm min}=0.03$ pc and $\lambda_{\rm max}=3.2$ pc. This length range corresponds to the core mass range $0.48 \ {\rm M_{\odot}} < M < 51.2 \  {\rm M_{\odot}}$. In our model we consider a uniform density filament, while density fluctuations are observed along filaments \citep{Roy2015}. Taking into account such density fluctuations would require a more detailed analysis, which are beyond the scope of this paper presenting first-order calculations toward understanding the angular momentum in the context of filament fragmentation.
\subsection{Velocity Field}
First we numerically generate the velocity field $\bi{v}$ in the filament following the method described in, e.g., \citet{Dubinski1995}
\begin{eqnarray}
\bi{v}(\bi{x})=\sum_{\bi{k}} \bi{V}(\bi{k}) \exp(\rmi \bi{k}\cdot \bi{x}),
\label{eq:velofieldfourier}
\end{eqnarray}
where $\bi{V}(\bi{k})$ is the Fourier transform, $\bi{k}$ is the wave vector. 
Since we consider only the periodic function for the velocity fluctuation, we include only the modes that satisfy the following condition:
\begin{eqnarray}
{\bi v}(x,y,z)={\bi v} (x+L_x,y,z)={\bi v}(x,y+L_y,z)={\bi v}(x,y,z+L_z),
\label{eq:veloperiod}
\end{eqnarray}
where we choose $L_x=L_y=2R_{\rm fil}$ for simplicity. $R_{\rm fil}$ is the radius of the filament, 0.05 pc in this paper. We define the power spectrum as
\begin{eqnarray}
\label{powerspedefinewq}
P(\bi{k}) = \left<|\bi{V}(\bi{k})|^2\right>,
\end{eqnarray}
where $\left< \ \right>$ represents the ensemble average. We describe in the following how we generate the Fourier component of $\bi{V}(\bi{k})$ for a given power spectrum $P(\bi{k})$ for both incompressible and compressible velocity fields.


\subsubsection{Incompressible Velocity Field}

For an incompressible fluid, $\bi{v}=\bi{\nabla} \times \bi{A}$, where $\bi{A}$ is the vector potential, 
\begin{eqnarray}
\bi{A}(\bi{x})=\sum_{\bi{k}} \bi{A}_{\bi{k}} \exp(\rmi \bi{k}\cdot \bi{x}),
\label{eq:vectorpotential1}
\end{eqnarray}
where $\bi{A}_{\bi{k}}$ is its Fourier transform. The Fourier component of the velocity field is
\begin{eqnarray}
\bi{V}(\bi{k}) = \rmi \bi{k} \times \bi{A}_{\bi{k}}.
\label{eq:vectorpotential2}
\end{eqnarray}
We generate the Fourier component of the vector potential $\bi{A}_{\bi k}$ as a random Gaussian number with a prescribed power spectrum, $P(\bi{k}) = (2/3)k^2 \left< |\bi{A}_{\bi k}|^2\right>$. Finally, we derive the velocity field in real space by performing the inverse transform of $\bi{V}({\bi k})$.\par
\subsubsection{Compressible Velocity Field}
We can also setup a compressible velocity field as, $\bi{v}=\bi{\nabla} \phi$, where $\phi$ is the scalar potential, 
\begin{eqnarray}
\phi(\bi{x})=\sum_{\bi{k}} \phi_{\bi{k}} \exp(\rmi \bi{k}\cdot \bi{x}),
\label{eq:vectorpotentialcom1}
\end{eqnarray}
and $\phi_{\bi{k}}$ is the Fourier transform. The Fourier component of the velocity field is
\begin{eqnarray}
\bi{V}(\bi{k}) = \rmi \bi{k} \phi_{\bi{k}}.
\label{eq:vectorpotentialcom2}
\end{eqnarray}
We generate the Fourier component of the scalar potential $\phi_{\bi k}$ as a random Gaussian number with a prescribed power spectrum, $P(\bi{k}) = k^2 \left< |\phi_{\bi k}|^2\right>$. Finally, we derive the velocity field in real space by performing the inverse transform of $\bi{V}({\bi k})$.\par


\subsection{Numerical Calculation of the Angular Momentum}
Since \citet{Inutsuka2001} has shown that the core mass function can be described by Press-Schechter formalism \citep{Press_Schechter1974}, we use the Press-Schechter formalism to derive the core mass and angular momentum of cores. In the Press-Schechter formalism, the length scale is defined as the collapsed region which would form a core in the future and the mass scale is defined as the mass in that collapsed region. Therefore, for a uniform filament, the mass is determined by the length (Figure \ref{Alfig1}). By assuming conservation of the angular momentum in that region, we adopt the angular momentum in that region as the angular momentum of the future core. \par
Following this concept, first, we choose an arbitrary length, $L_{\rm core}$, at a random position along the longitudinal direction of the filament. Next, we calculate the angular momentum in that region as follows:
\begin{eqnarray}
\label{ang}
\bi{J}(M) = \rho \int \bi{x} \times \bi{v} \ d^3x,
\end{eqnarray}
where $\rho$ is the density, $\bi{x}$ is the position vector. We repeat this procedure for 99 values of $L_{\rm core}$ between $\lambda_{\rm min}$ and $\lambda_{\rm max}$ to obtain $j-M$ relation. By using this method, we can study how the $j-M$ relation depends on the power spectrum of the velocity field.\par
In this work we examine four power spectrum models: a three-dimensional (3D) Kolmogorov power spectrum (Section 3.1), a log-normal power spectrum (Section 3.2), an anisotropic power spectrum (Section 3.3), and a one-dimensional (1D) Kolmogorov power spectrum (Section 3.4).


\subsection{Analytical Solution for Isotropic Velocity Field}
For an isotropic velocity field, we can analytically derive the angular momentum as follows. 


\subsubsection{Incompressible Velocity Field}
The angular momentum in the region with a length scale $L_{\rm core}=M/{\rm M}_{\rm line,crit}$ is given by
\begin{eqnarray}
\label{ceq1}
\bi{J}(M) & = \rho \int \bi{x} \times \bi{v} d^3x \nonumber \\
&= - \rho \sum_{\bi{k}} \bi{V}(\bi{k}) \times \int \bi{x} \exp(\rmi \bi{k}\cdot \bi{x}) d^3x\nonumber \\
&= 2 \rmi M\sum_{\bi{k}} \bi{V}(\bi{k}) \times  \bi{R}(\bi{k};L_{\rm core}).
\end{eqnarray}
$\bi{R}(\bi{k};L_{\rm core})$ is defined as
\begin{eqnarray}
\label{Rdef}
\bi{R}(\bi{k};L_{\rm core}) \equiv \frac{\rmi}{2\pi R_{\rm fil}^2 L_{\rm core}} \int \bi{x} \exp(\rmi \bi{k}\cdot \bi{x})d^3x.
\end{eqnarray}
$\bi{R}(\bi{k};L_{\rm core})$ corresponds to the Fourier transform of the position vector $\bi{x}$. The detailed derivation of $\bi{R}(\bi{k};L_{\rm core})$ is shown in Appendix A.
Using \Eref{ceq1}, we can derive the specific angular momentum
\begin{eqnarray}
\label{ceq10}
j(M) &= \frac{\sqrt{\left<J(M)^2 \right>}}{M} \nonumber \\
&=  \sqrt{2 \sum_{\bi{k}}P(k) \left[\{\bi{R}(\bi{k})\}^2  +  \frac{ \{\bi{k}\cdot\bi{R}(\bi{k}) \}^2 }{k^2} \right]},
\end{eqnarray}
where $P(k)$ is the power spectrum as introduced in Section 2.3. $j^2$ is the sum of the product of the square of the Fourier components of the velocity ($P(k) = \left<|\bi{V}(\bi{k})|^2\right>$) and the position ($\bi{R}(\bi{k};L_{\rm core})^2$).

\subsubsection{Compressible Velocity Field}
In the case of the potential velocity field, we can also derive an analytical expression for the specific angular momentum similar to \Eref{ceq10}. The specific angular momentum is
\begin{eqnarray}
\label{ceq13}
j(M) &= \frac{\sqrt{\left<J(M)^2\right>}}{M} \nonumber \\
&= 2 \sqrt{ \sum_{\bi{k}} P_{\phi}(k)[ \bi{k}^2 \{\bi{R}(\bi{k})\}^2 - \{\bi{k}\cdot \bi{R}(\bi{k}) \}^2 ] },
\end{eqnarray}
where $P_{\phi}$ is defined as
\begin{eqnarray}
\label{ceq14}
P_{\phi}(k) = \left<|\phi_{\bi{k}}|^2 \right>.
\end{eqnarray}


\section{Results}
In this section, we examine four power spectrum models: three isotropic models (3D Kolmogorov power spectrum model in Section 3.1, log-normal power spectrum model in Section 3.2, and 1D Kolmogorov power spectrum model in Section 3.4) and one anisotropic power spectrum model in Section 3.3. We compare the $j-M$ relations derived from our four models over the mass range $0.48 \ {\rm M_{\odot}} < M < 51.2 \ {\rm M_{\odot}}$ (cf., Section 2.1) with the observational results to discuss the origin of the observed angular momentum of cores. 


\subsection{3D Kolmogorov Power Spectrum Model}
In this subsection, we adopt a 3D Kolmogorov power spectrum compatible with the observed Kolmogorov turbulent power spectrum of the ISM {\citep{Armstrong1995}, 
\begin{eqnarray}
\label{eq:3DKPS}
P(k)dk_x dk_y dk_z =  Ak^{-11/3}dk_x dk_y dk_z,
\end{eqnarray}
where $k=\sqrt{k_x^2+k_y^2+k_z^2}$. Note, however, that hereafter we consider only the discrete modes that are periodic in the domain ($-L_x/2 < x < L_x/2, -L_y/2 < y < L_y/2, -L_z/2 < z < L_z/2$) as in \Eref{eq:veloperiod}. 
If we define the major axis of the filament as the $z$-axis ($x=y=0$), we can define the velocity along the filament as follows,
\begin{eqnarray}
\label{eq:3DKPSalong}
{\bi v}_{\rm 1D}(z) = \frac{1}{M_{\rm line}}\int \int^{\sqrt{x^2+y^2}<R_{\rm fil}} \rho {\bi v} \ dx dy,
\end{eqnarray}
where $R_{\rm fil}$ is the radius of the filament. If we use ${\bi v}$ following to the power spectrum \Eref{eq:3DKPS}, the slope of the power spectrum of ${\bi v}_{\rm 1D}(z)$ is $-11/3$ (cf., Section 5).
The coefficient A reflects the 3D velocity dispersion, $\sigma_{\rm 3D}$, along the filament crest. The observed velocity dispersion towards filaments with ${\rm M}_{\rm line} \sim {\rm M}_{\rm line,crit}$ is $c_s \lesssim \sigma_{\rm 1D} \lesssim 2c_s$ \citep{Hacar2011,Arzoumanian2013,Hacar2016} with $\sigma_{\rm 1D}=\sigma_{\rm 3D}/\sqrt{3}$. Since $\sigma_{\rm 3D}$ and $j$ are proportional to $\sqrt{A}$, setting $\sigma_{\rm 3D}$ between $c_s$ and $2c_s$ will change the result by a factor of only 2. We therefore choose $\sigma_{\rm 3D}=c_s$ as our fiducial value for the models and for the figures shown in this paper. In Section 4 we compare the results obtained with $\sigma_{\rm 3D}=c_{\rm s}$ and $\sigma_{\rm 3D}=2c_{\rm s}$.\par
Figure \ref{CAcfullkolonly} shows the $j-M$ relation obtained from the 3D Kolmogorov power spectrum model. The red filled circles and the solid line represent the numerical and the analytical results, respectively. 
\begin{figure}[t]
\centering
\includegraphics[width=10cm]{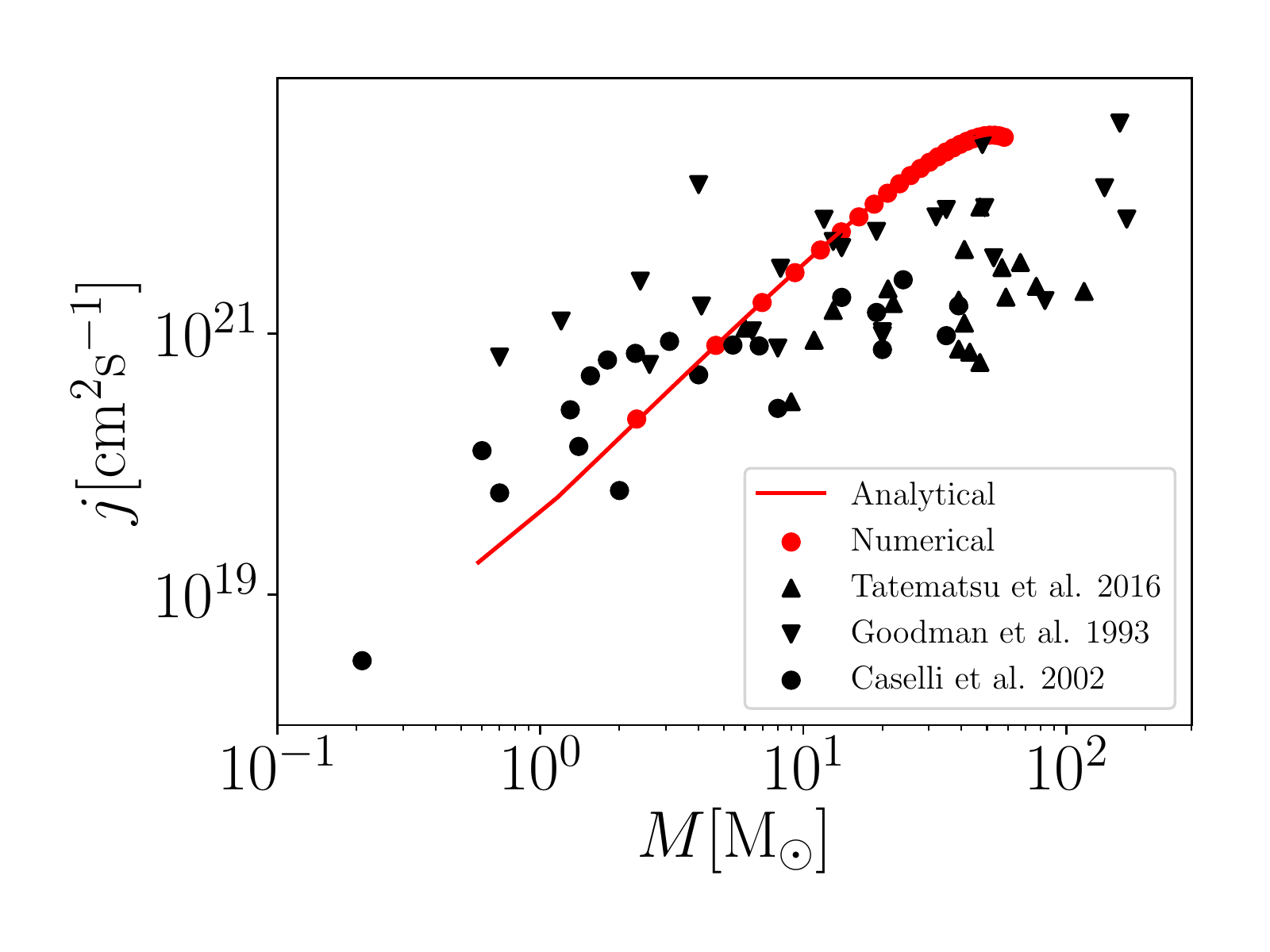}
\caption{$j-M$ relation obtained from 3D Kolmogorov turbulence power spectrum model (Section 3.1). The red filled circles and the solid line represent the numerical and the analytical results, respectively. The observational data are the same as in Figure \ref{goodmancasellitatematsu}.}
\label{CAcfullkolonly}
\end{figure}
The analytical result is calculated using \Eref{ceq10}. Since, as can be seen in Figure \ref{CAcfullkolonly}, the analytical result (\Eref{ceq10}) agrees with the numerical result, in the following Section 3.2 and Section 3.4 we will continue the discussion using the analytical calculation.\par
Figure \ref{CAcfullkolonly} suggests that for the 3D Kolmogorov power spectrum, the observed $j-M$ relation is not reproduced. This result can be understood as follows. The velocity fluctuations along a filament correspond to a superposition of different modes with different wavenumbers. The wavenumber which is the most important for the angular momentum in the region with a length $L_{\rm core}$ is $k=2\pi /(2L_{\rm core})$. The modes with large wavenumbers do not contribute to the integration of the right hand side of \Eref{ang} since these modes cancel out (Figure \ref{ameffectivewavepaper}). Hence the small angular momentum derived with this model for low mass regions is due to the small amount of energy in small wavelength (large wavenumber) regions. The large angular momentum derived for the high mass region is also inconsistent with the observations (Figure \ref{CAcfullkolonly}). In the following section we address this problem. 
\begin{figure}[t]
\begin{tabular}{cc}
\begin{minipage}[t]{.35\textwidth}
\centering
\includegraphics[width=7cm]{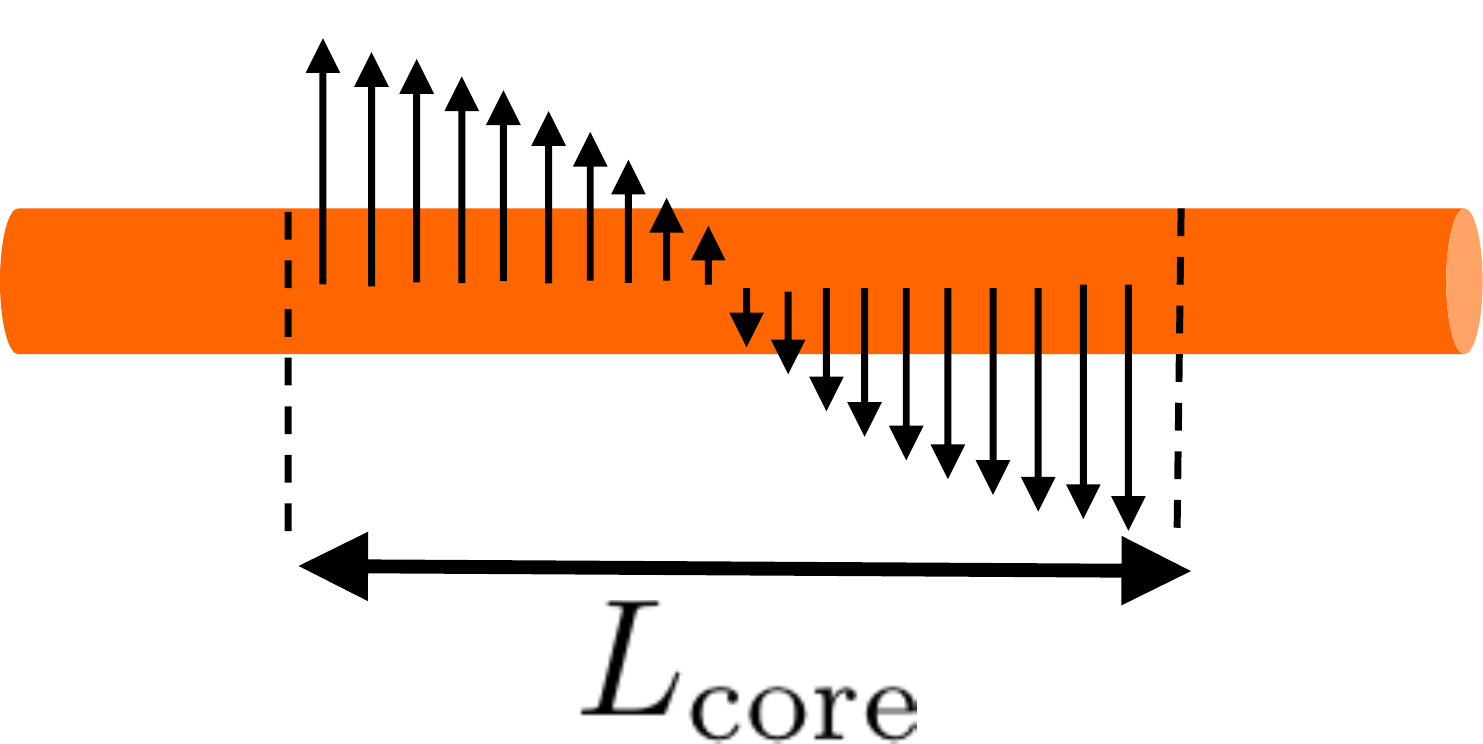}
\end{minipage}
\begin{minipage}{.15\textwidth}
\hspace{10mm}
\end{minipage}

\begin{minipage}[t]{.35\textwidth}
\centering
\includegraphics[width=7cm]{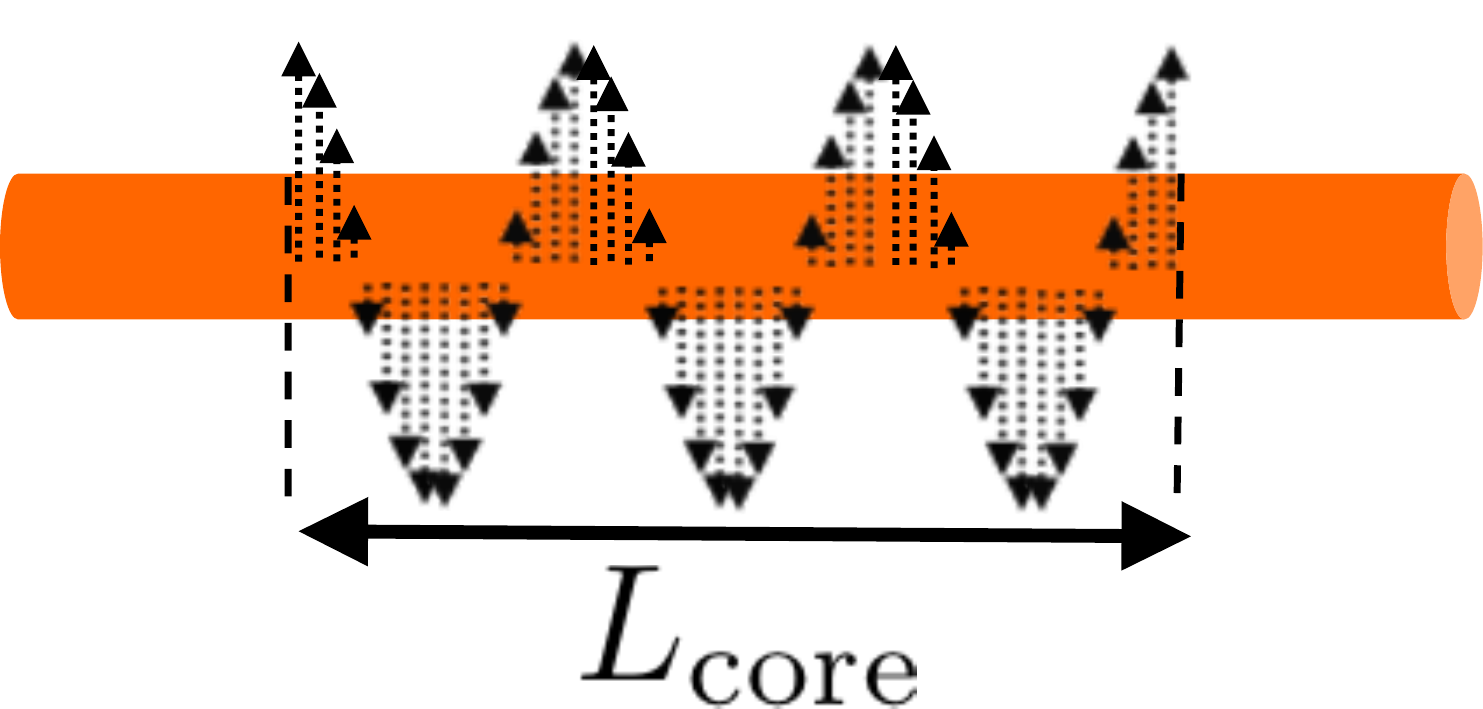}
\end{minipage}
\end{tabular}
\caption{Example of the superposition of different fluctuating modes along the filament. In the region with a length $L_{\rm core}$ the wave represented by the solid line arrows is the most important for the angular momentum (left) and the wave represented by the dashed line arrows mostly cancels and hence does not significantly contribute to the angular momentum of this region (right).}
\label{ameffectivewavepaper}
\end{figure}


\subsection{Log-normal Power Spectrum Model}
To analyze the problem mentioned in the previous subsection, one may increase the power in large wavenumber region compared to the 3D Kolmogorov power spectrum. To do so, in this subsection, we adopt a log-normal energy spectrum,   
\begin{eqnarray}
E(k) = A_G \exp \left[ -\frac{\left\{\log(k) - \log(k_{\rm peak})\right\}^2}{2\sigma_G^2} \right],
\label{eq:PSgauss}
\end{eqnarray}
where $A_G$, $\sigma_G$, and $k_{\rm peak}$ are the amplitude, the dispersion, and the peak of the log-normal function, respectively. The energy spectrum is defined as follows,
\begin{eqnarray}
\frac{\left<|\bi{v}|^2 \right>}{2} = \sum_{\bi k} E(k).
\label{eq:defEk}
\end{eqnarray}
The power spectrum is described as 
\begin{eqnarray}
P(k)dk_x dk_y dk_z = \frac{A_G}{2\pi k^2} \exp \left[ -\frac{\left\{\log(k) - \log(k_{\rm peak})\right\}^2}{2\sigma_G^2} \right]dk_x dk_y dk_z.
\label{eq:PSgausspower}
\end{eqnarray}
As in Section 3.1, we consider only the discrete modes that are periodic in the domain.


\subsubsection{Log-normal Power Spectrum Model for Incompressible Velocity Field}
We regard $k_{\rm peak}$ and $\sigma_G$ in \Eref{eq:PSgausspower} as parameters to study how $j-M$ relation depends on the shape of the energy spectrum, and choose $A_G$ to satisfy the constraint $\sigma_{\rm 3D} = c_s$. Figure \ref{CALogGaussiansigmacompare} shows the $j-M$ relation derived from \Eref{ceq10} and \Eref{eq:PSgausspower}. Here we use $k_{\rm peak}=2\pi / (0.1 \  {\rm pc})$ for a filament width of 0.1 pc, and regard only $\sigma_G$ as a variable parameter ($\sigma_G =$ 0.35, 0.5, and 0.65). Figure \ref{CALogGaussianpeakcompare} shows the $j-M$ relation for fixed $\sigma_G =$ 0.5 and varying $\lambda_{\rm peak}=$ 1.0, 0.5, 0.1, and 0.05 pc, respectively with $\lambda_{\rm peak}$ defined as $k_{\rm peak} = 2\pi /\lambda_{\rm peak}$. In this model the power of the wave with wavelength $\lambda_{\rm peak}$ is the largest. Since, in our calculation, the waves along $x$- or $y$-axis with wavelength larger than 0.1 pc are truncated, their component along the $z$-axis has the largest power in the models with $\lambda_{\rm peak}>0.1$ pc. The results shown in Figure \ref{CALogGaussiansigmacompare} and Figure \ref{CALogGaussianpeakcompare} indicate that the power spectrum described by \Eref{eq:PSgausspower} reproduces the observed trend of the angular momentum of cores. Figure \ref{kavesigmagindex} shows how the index depends on the two parameters $k_{\rm peak}$ and $\sigma_G$. Since the $j-M$ relations derived from the log-normal model are not straight lines, we apply the least square fitting method to the derived $j-M$ curves in the mass range, $1-10 \ {\rm M}_{\odot}$ to derive the power law index of each model curve. While the comparison with observations suggests that the index is approximately $0.5-0.9$, future observations are needed to better constrain this index value. Once the $j-M$ relation for the angular momentum distribution is accurately determined by observations, we would be able to give, using our model, a more accurate description of the velocity fluctuations along the parent filament.
\begin{figure}[t]
\begin{tabular}{cc}
\begin{minipage}[t]{.5\textwidth}
\centering
\includegraphics[width=8cm]{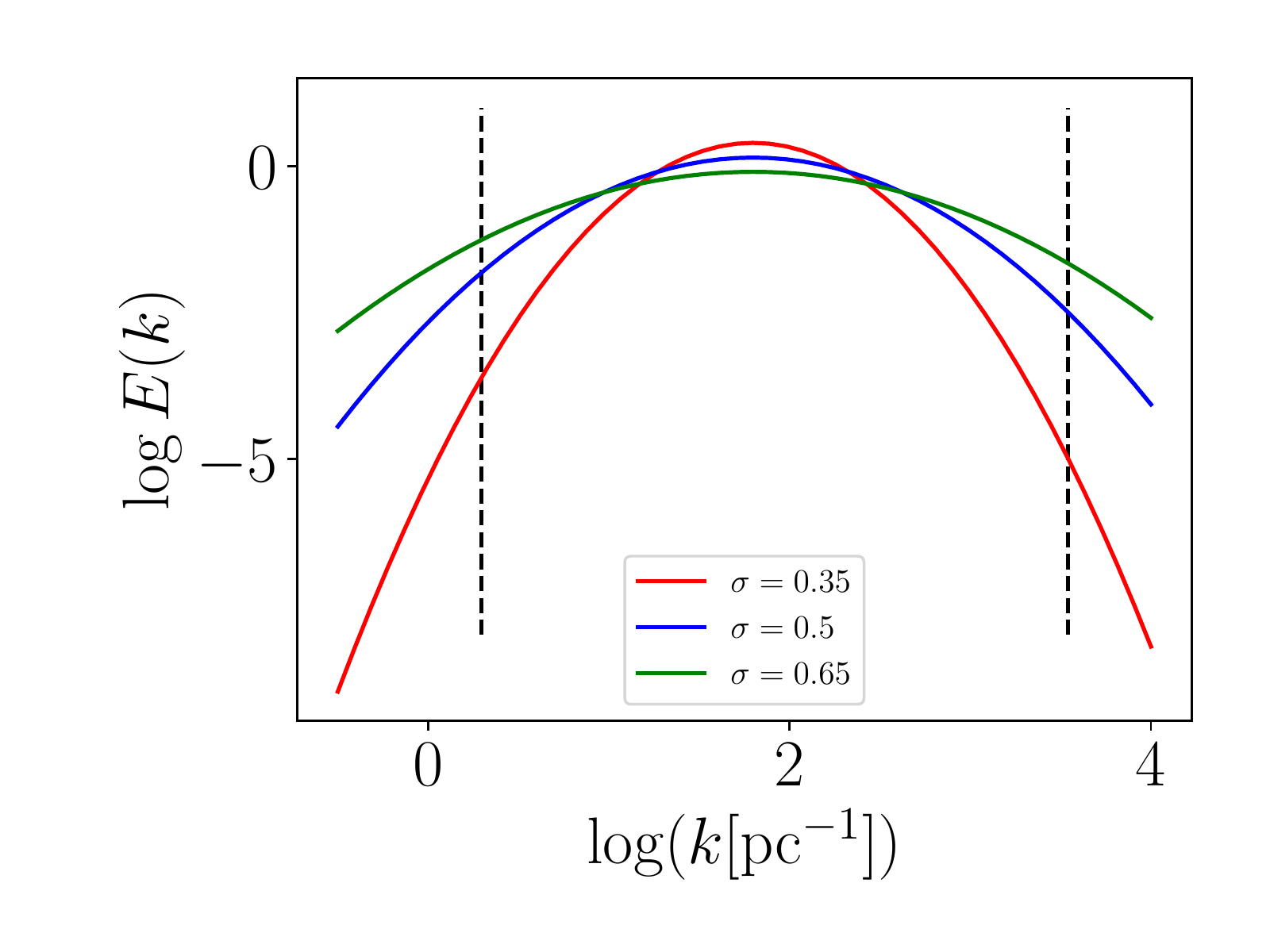}
\end{minipage}
\begin{minipage}[t]{.5\textwidth}
\centering
\includegraphics[width=8cm]{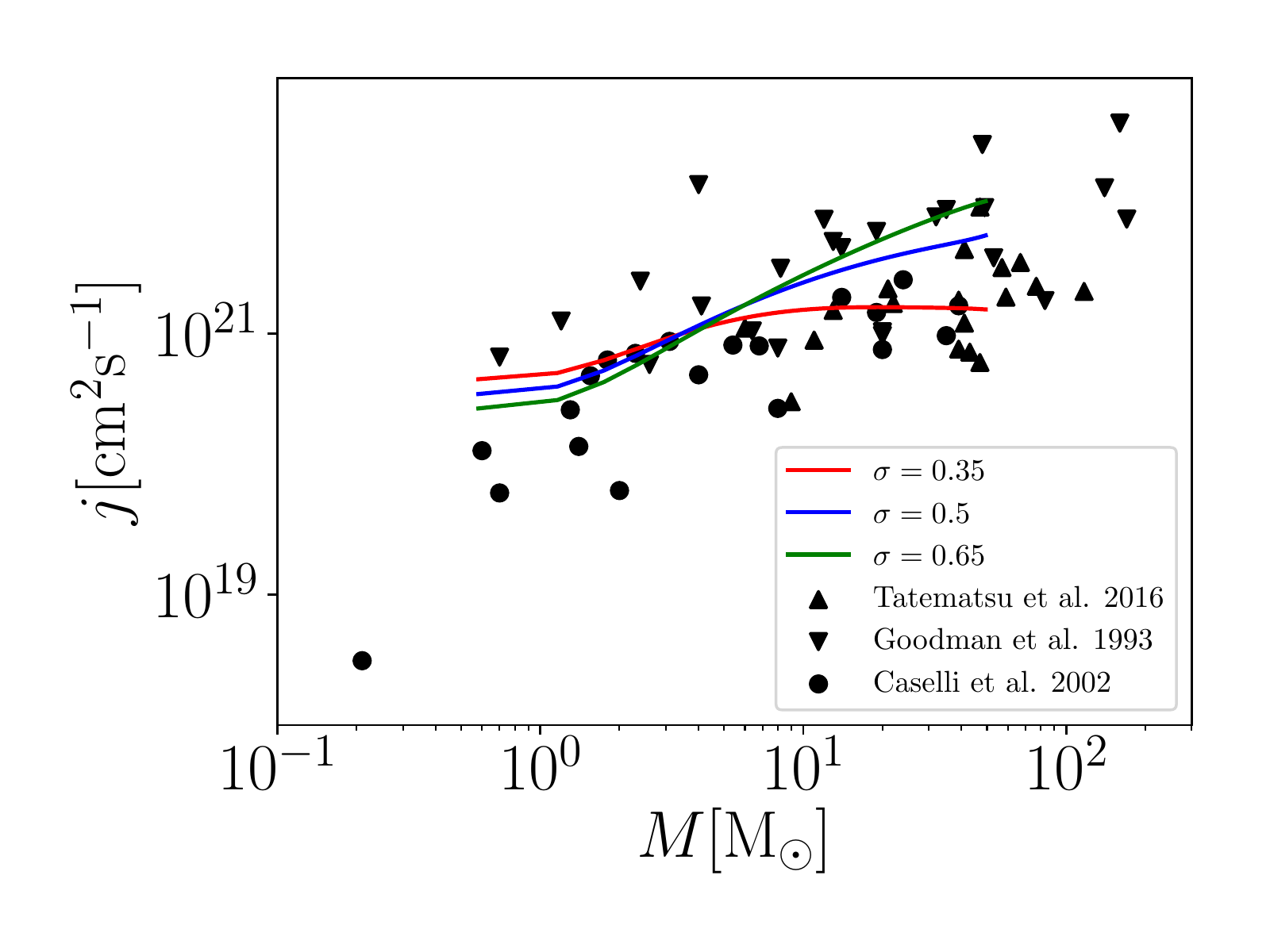}
\end{minipage}
\end{tabular}
\caption{(left) The horizontal axis is the wavenumber and the vertical axis is the energy spectrum. The region between the two black dashed lines represents the range of wavenumber that we used. (right) $j-M$ relation derived from the log-normal power spectrum model with fixed value, $k_{\rm peak}=2\pi / (0.1 {\rm pc})$. The red, blue, and green solid lines correspond to $\sigma_G = 0.35, 0.5$, and 0.65, respectively. The observational data are the same as in Figure \ref{goodmancasellitatematsu}.}
\label{CALogGaussiansigmacompare}
\end{figure}

\begin{figure}[t]
\begin{tabular}{cc}
\begin{minipage}[t]{.5\textwidth}
\centering
\includegraphics[width=8cm]{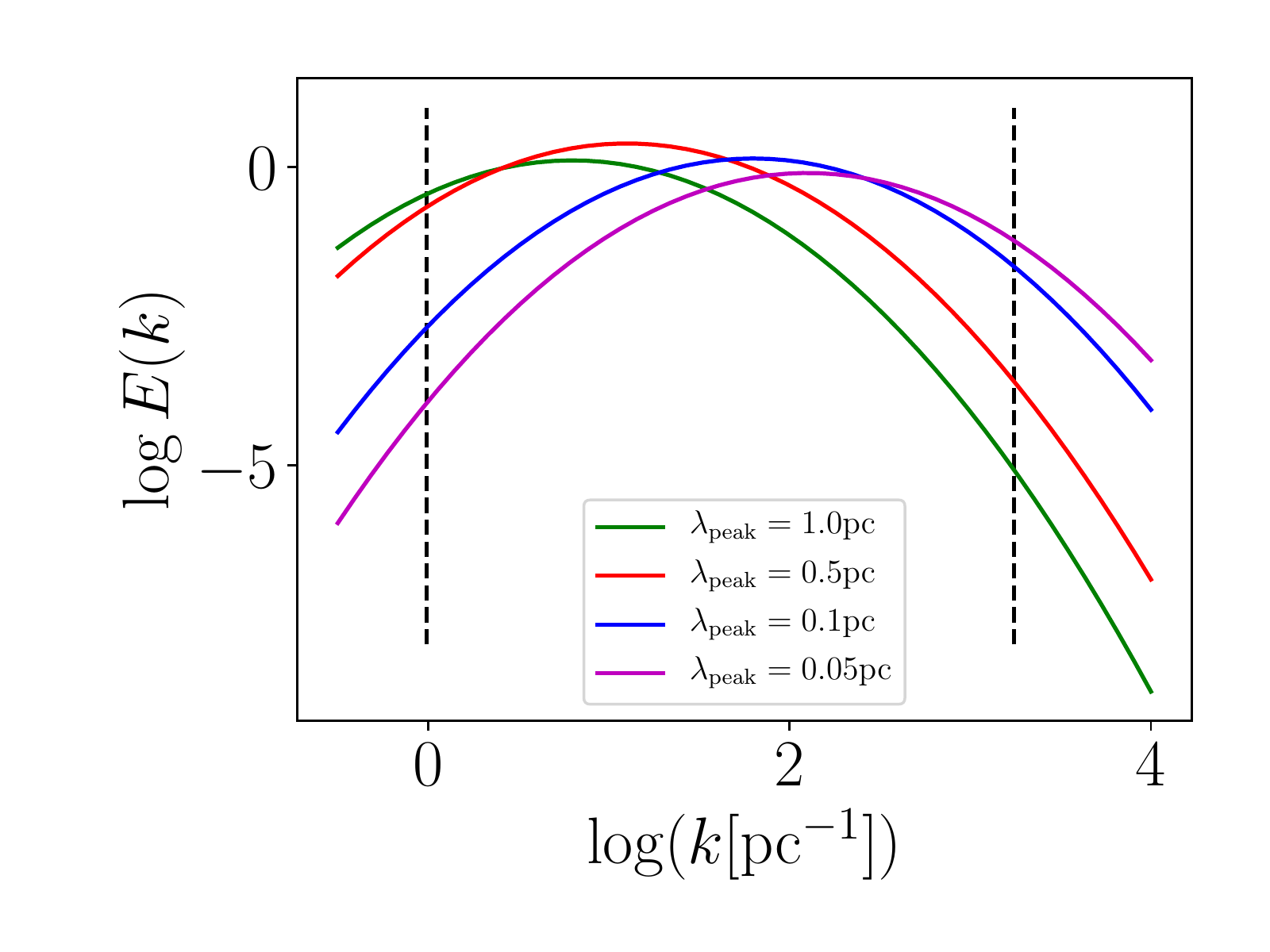}
\end{minipage}
\begin{minipage}[t]{.5\textwidth}
\centering
\includegraphics[width=8cm]{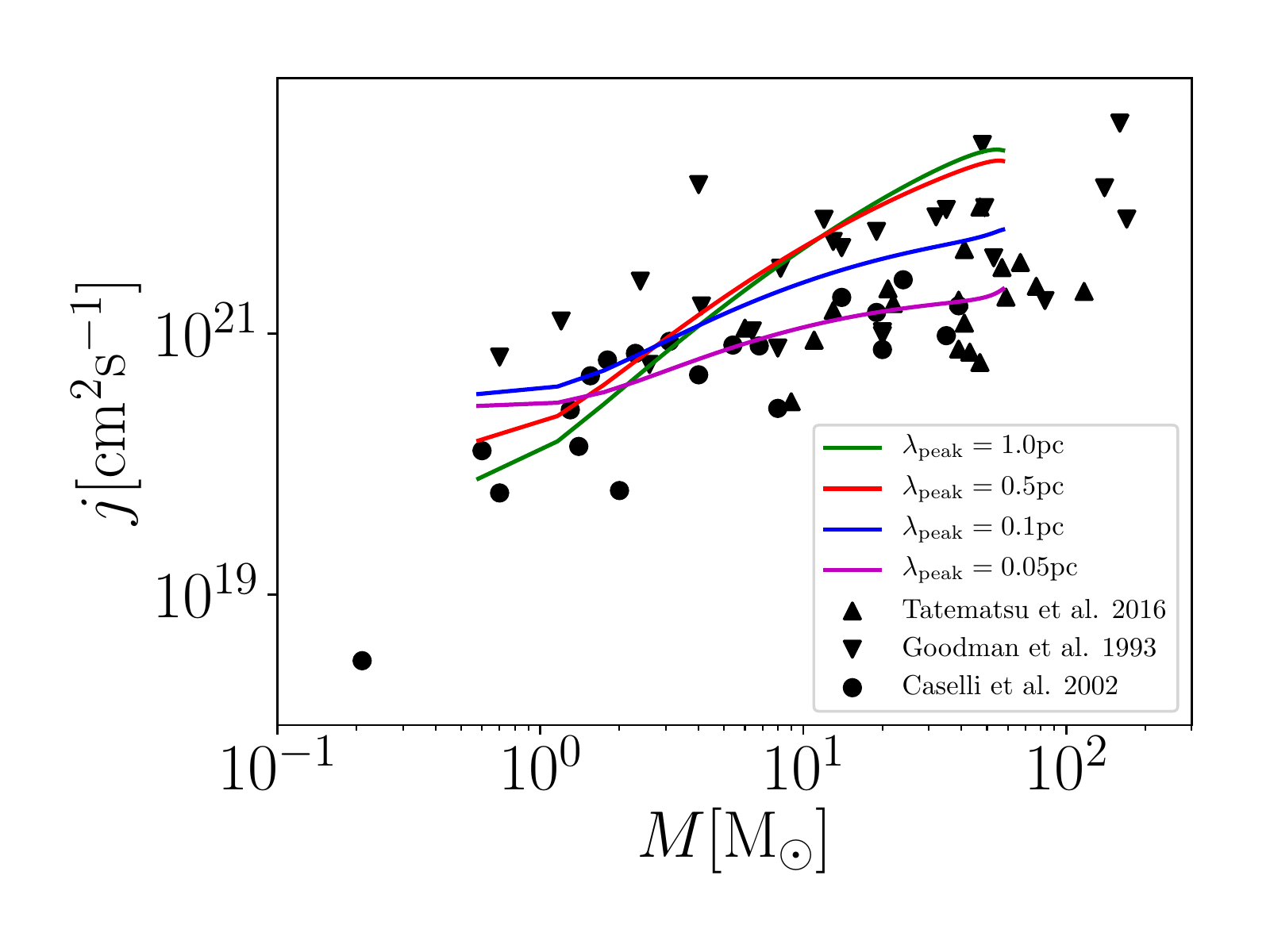}
\end{minipage}
\end{tabular}
\caption{(left) Same as Figure \ref{CALogGaussiansigmacompare} for fixed $\sigma_G = 0.5$ and varying $\lambda_{\rm peak}$. (right) $j-M$ relation derived from the log-normal power spectrum model with fixed value $\sigma_G=0.5$. The green, red, blue, and magenta solid lines correspond to $\lambda_{\rm peak}=1.0,0.5,0.1$, and 0.05 pc. $\lambda_{\rm peak}$ is defined as $k_{\rm peak} = 2\pi /\lambda_{\rm peak}$. The observational data are the same as in Figure \ref{goodmancasellitatematsu}.}
\label{CALogGaussianpeakcompare}
\end{figure}

\begin{figure}[t]
\centering
\includegraphics[width=10cm]{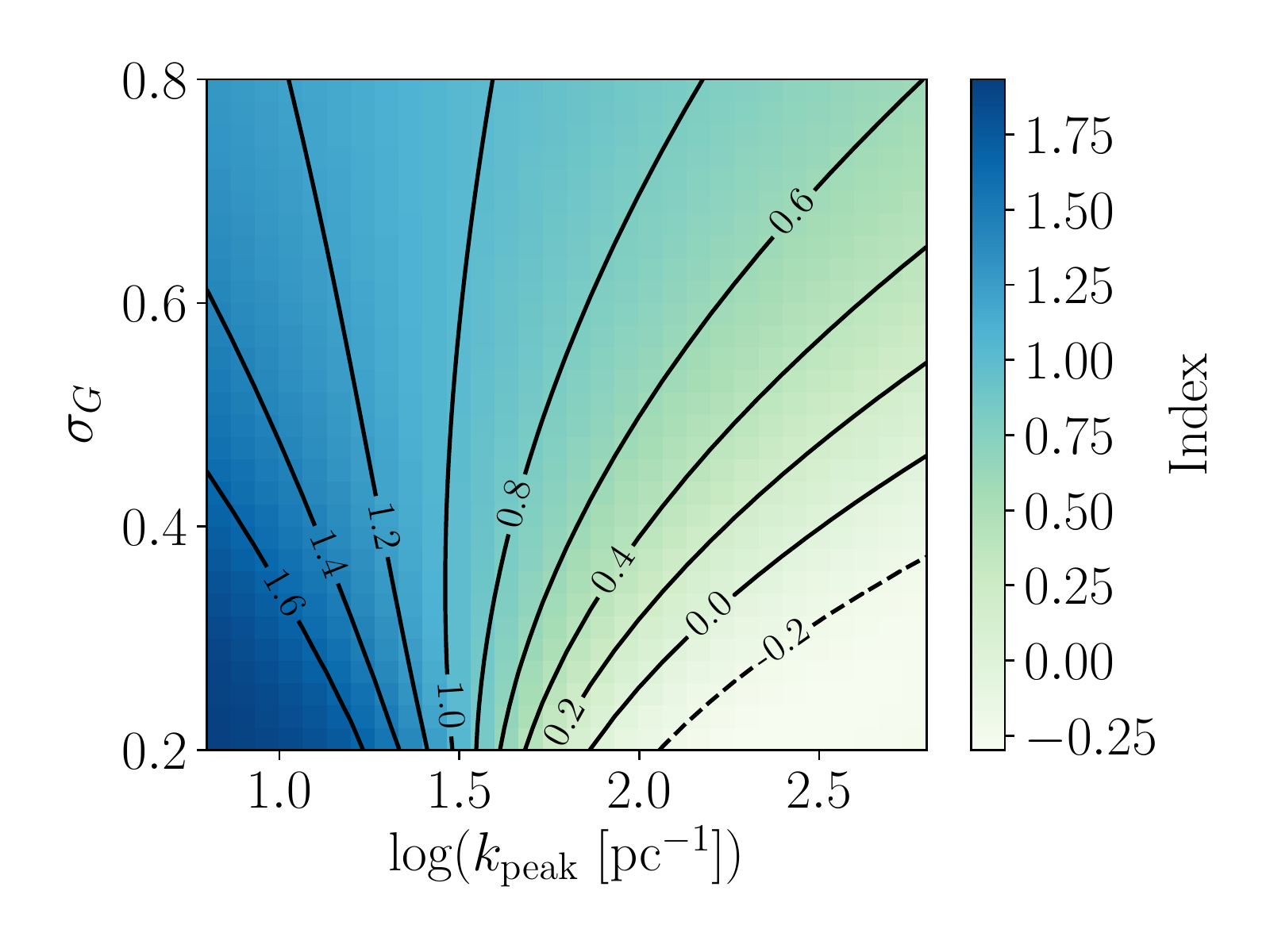}
\caption{Dependence of the index of the $j-M$ relation slope on the two parameters, $\sigma_G$ and $k_{\rm peak}$. The horizontal axis is $k_{\rm peak}$ and the vertical axis is $\sigma_G$. The indices are derived by applying the least square fitting method to the $j-M$ relation in the mass range $1-10\ {\rm M}_{\odot}$. }
\label{kavesigmagindex}
\end{figure}


\subsubsection{Log-normal Power Spectrum Model for Compressible Velocity Field}
\begin{figure}[t]
\centering
\includegraphics[width=10cm]{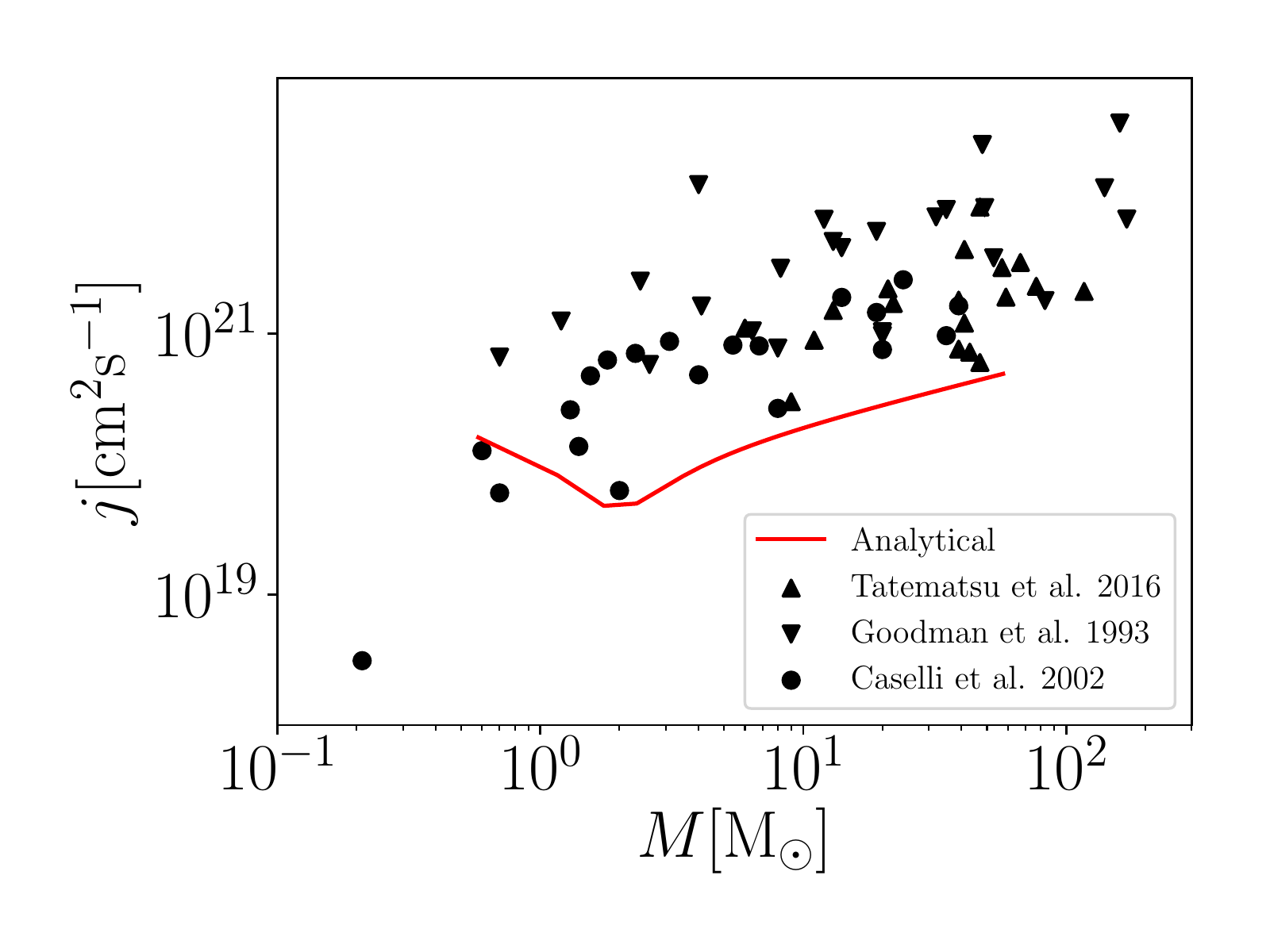}
\caption{$j-M$ relation for a compressible velocity field. The red solid line is the result obtained from the expression for the potential field shown in \Eref{ceq13}, and the log-normal power spectrum \Eref{eq:PSgausspower} with $\sigma_G=0.5$ and $\lambda_{\rm peak}=0.1$ pc. The observational data are the same as in Figure \ref{goodmancasellitatematsu}.}
\label{CAcsolenoidalflatmode01pcc}
\end{figure}
We also calculate the angular momentum in the case of compressible velocity field using \Eref{ceq13} for $\sigma_{\rm 3D} = c_s$. Figure \ref{CAcsolenoidalflatmode01pcc} shows the $j-M$ relation for a potential velocity field using the log-normal power spectrum model described by \Eref{eq:PSgausspower}, for $\sigma_G=0.5$ and $\lambda_{\rm peak}=0.1$ pc. As shown in Figure \ref{CAcsolenoidalflatmode01pcc}, the compressible velocity field hardly contributes to the angular momentum of cores. This result suggests that the solenoidal component of the velocity field in filaments mainly accounts for the origin of the angular momentum of the molecular cloud cores. Therefore, hereafter we consider only the incompressible velocity component. 


\subsection{Anisotropic Power Spectrum Model}
As mentioned in Section 3.2.1, with an isotropic power spectrum (\Eref{eq:PSgauss}), we can reproduce the properties of the angular momentum of cores with the available observed data points. However, such an increase in the power of velocity or density fluctuations in larger wavenumber (small scales) has not been observationally reported, possibly because of the limited spatial resolution \citep{Roy2015}. In contrast, for the density distribution in the larger wavelength range ($>$0.02 pc), relatively simple Kolmogorov-like power laws are observed along filaments and in the diffuse ISM \citep{Roy2015,Miville2010,Roy2019}. According to the recently proposed scenario of filamentary structure formation where filaments may be formed by large scale compressions \citep[e.g.,][]{InoueInutsuka2012,Inutsuka2015,Inoue2017,Arzoumanian2018}, it is expected that the waves along the $x$- and $y$-axis (transverse direction) have more energy than those along $z$-axis (longitudinal direction) as a result of energy shift from large to small scales due to the compression. In addition, transverse velocity gradients are reported by recent observations in several molecular filaments \citep[e.g.,][]{FernandezLopez2014,Dhabal2018}. Hence it is interesting to examine whether anisotropic power spectrum reproduces the observed $j-M$ relation. In this paper, for simplicity, we adopt the following simple Kolmogorov-like anisotropic power spectrum
\begin{eqnarray}
P(\bi{k})dk_x dk_y dk_z = A(A_r k_r^2 + k_z^2)^{-11/6}dk_x dk_y dk_z,
\label{eq:anisops}
\end{eqnarray}
where $k_r=\sqrt{k_x^2+k_y^2}$, and $A_r$ corresponds to the degree of anisotropy. As in Section 3.1 and 3.2, we consider only the discrete modes that are periodic in the domain.
\begin{figure}[t]
\begin{tabular}{cc}
\begin{minipage}[t]{.5\textwidth}
\centering
\includegraphics[width=8cm]{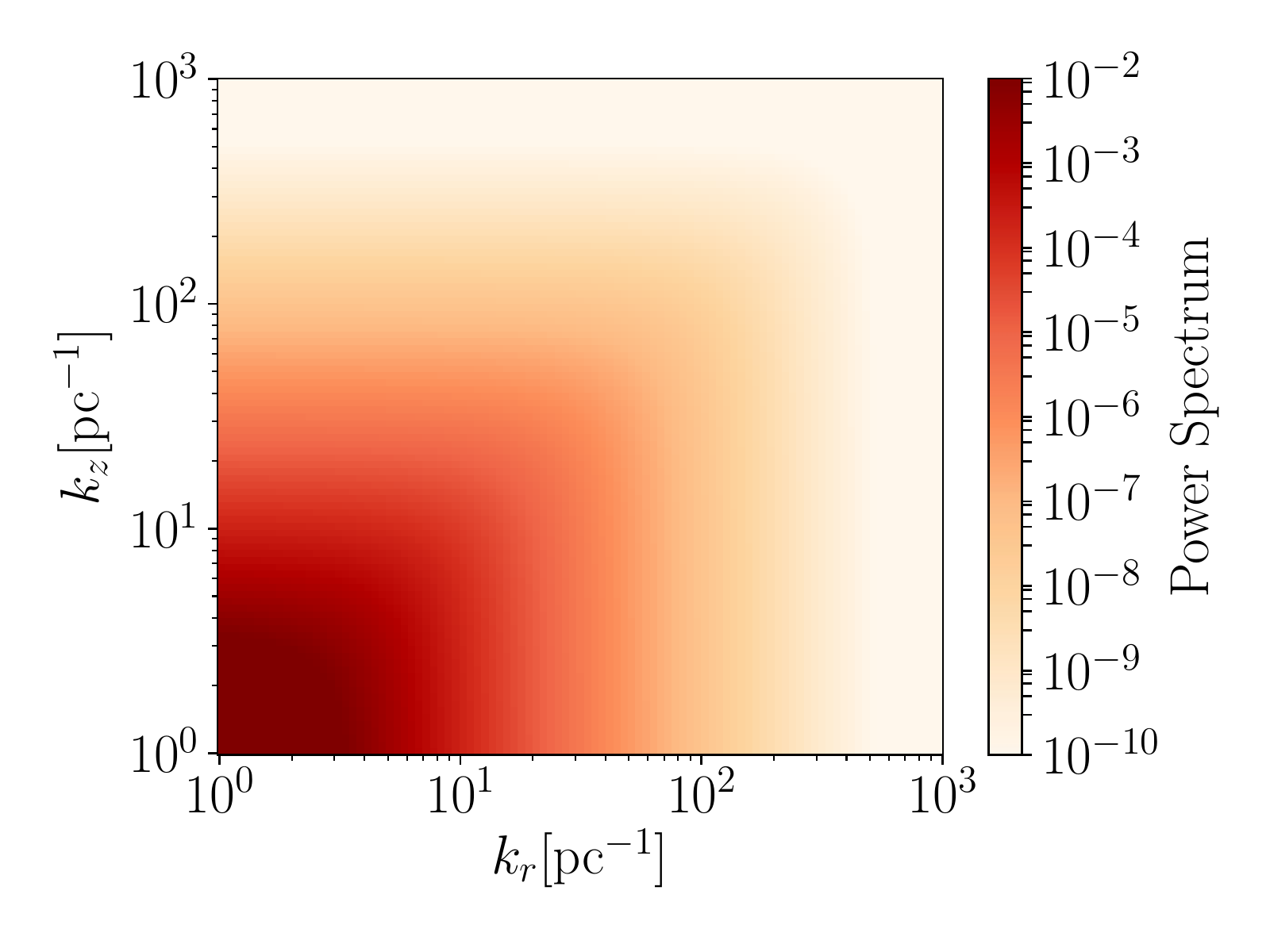}
\end{minipage}
\begin{minipage}[t]{.5\textwidth}
\centering
\includegraphics[width=8cm]{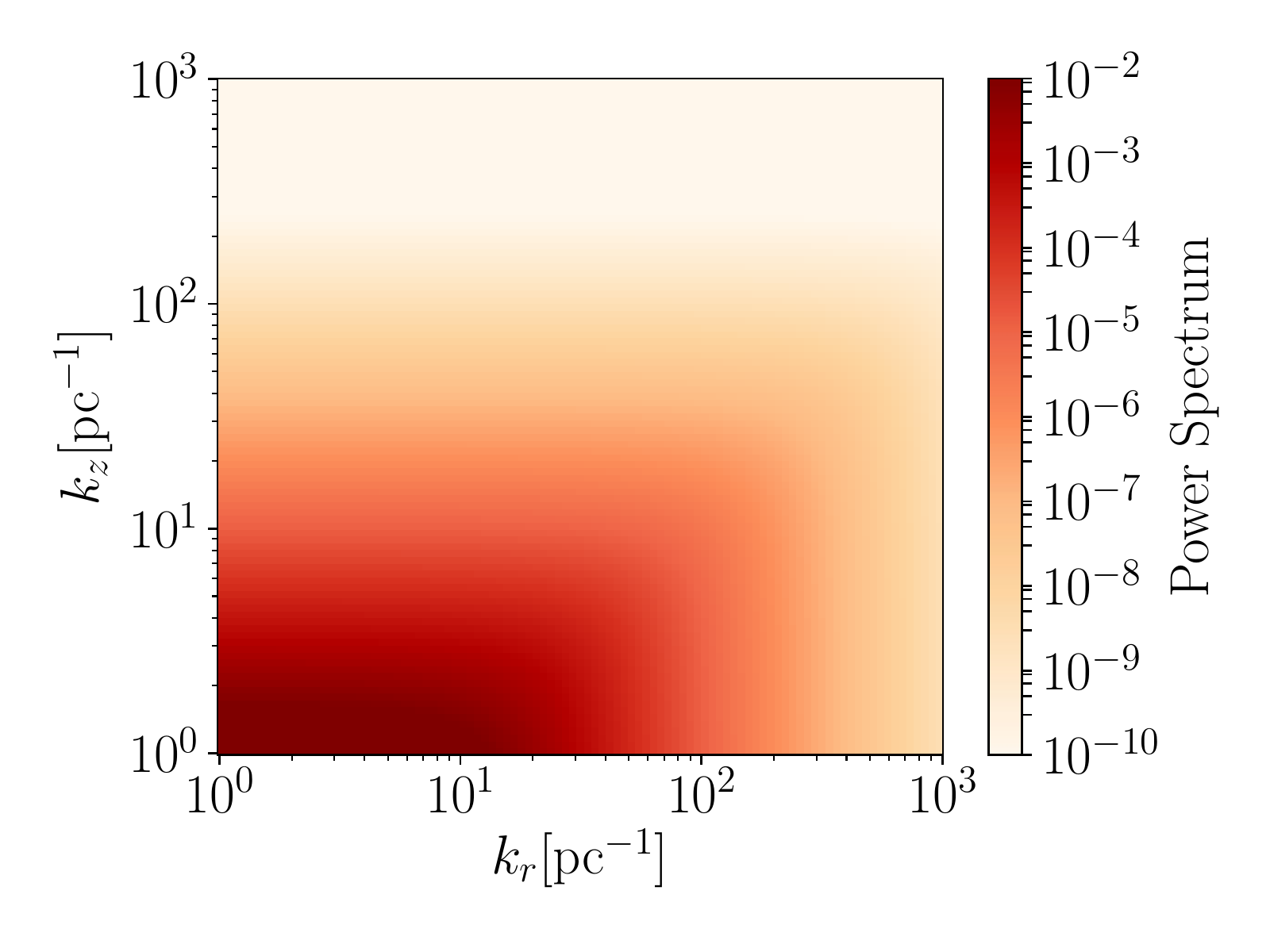}
\end{minipage}
\end{tabular}
\caption{Color contour of power spectrum. Isotropic 3D Kolmogorov power spectrum (left) and anisotropic model derived from \Eref{eq:anisops} with $A_r=0.01$ (right).}
\label{anisopow}
\end{figure}
Figure \ref{anisopow} shows the color contour for isotropic and anisotropic power spectra. Figure \ref{CAcfanisotropiccompareletter} shows the result from the anisotropic power spectrum with $A_r=0.01$ as shown in Figure \ref{anisopow} (right) with more energy in the transverse direction than in the longitudinal direction. We find that such an anisotropic power spectrum reproduces the observed properties of angular momentum of cores.
\begin{figure}[t]
\centering
\includegraphics[width=10cm]{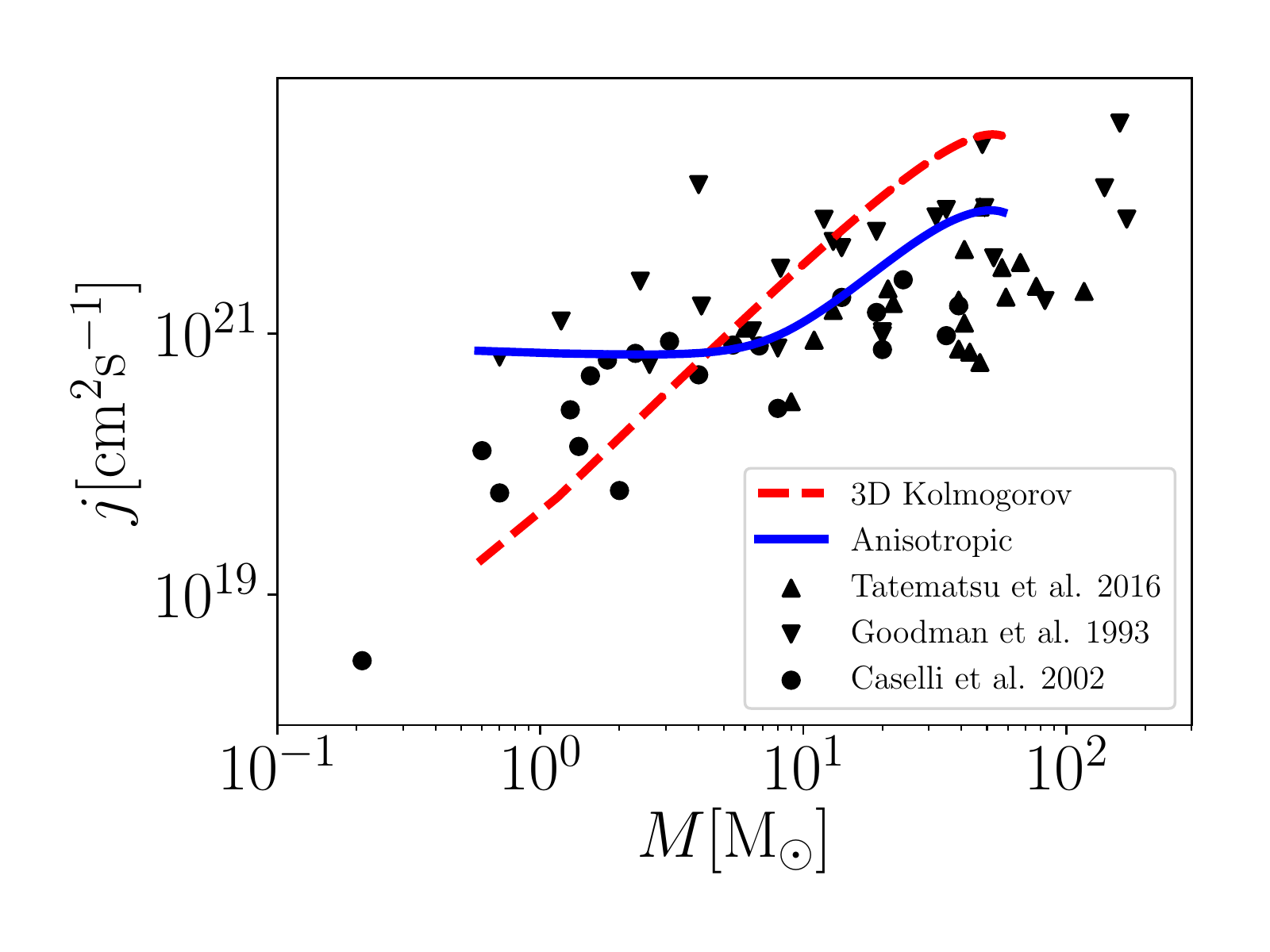}
\caption{$j-M$ relation for anisotropic power spectrum. The red dashed line and blue solid line are isotropic 3D Kolmogorov power spectrum (same as Figure \ref{CAcfullkolonly}) and anisotropic power spectrum model derived using \Eref{eq:anisops}, $A_r=0.01$ respectively. The observational data are the same as in Figure \ref{goodmancasellitatematsu}.}
\label{CAcfanisotropiccompareletter}
\end{figure}


\subsection{1D Kolmogorov Power Spectrum Model}
In this subsection, we examine the 1D Kolmogorov power spectrum model. The power spectrum is described as
\begin{eqnarray}
P(k)dk_x dk_y dk_z = Ak^{-5/3}dk_x dk_y dk_z,
\label{eq:1dkol}
\end{eqnarray}
where $k=\sqrt{k_x^2+k_y^2+k_z^2}$. We consider only the discrete modes that are periodic in the domain again. We can calculate the velocity along the filament using \Eref{eq:3DKPSalong}. If we use ${\bi v}$ following the power spectrum \Eref{eq:1dkol}, the slope of the power spectrum of ${\bi v}_{\rm 1D}(z)$ is $-5/3$ (cf., Section 5). The coefficient $A$ is chosen to satisfy the constraint $\sigma_{\rm 3D} = c_s$. Since the slope of the 1D Kolmogorov power spectrum is shallower than that of the 3D Kolmogorov power spectrum, the power in large wavenumber region is larger than for the 3D Kolmogorov power spectrum. Figure \ref{AMjM5over3} shows the $j-M$ relation derived from the 1D Kolmogorov power spectrum. We find that such a 1D Kolmogorov power spectrum also reproduces the observed properties of angular momentum of cores. \par
We think that the power spectrum of the filament should be determined in the formation process of the filament and the resultant power spectrum should be related to the 3D power spectrum in the parent molecular cloud. We discuss possible implications from our results in the formation of filamentary molecular clouds in Section 5.

\begin{figure}[t]
\centering
\includegraphics[width=10cm]{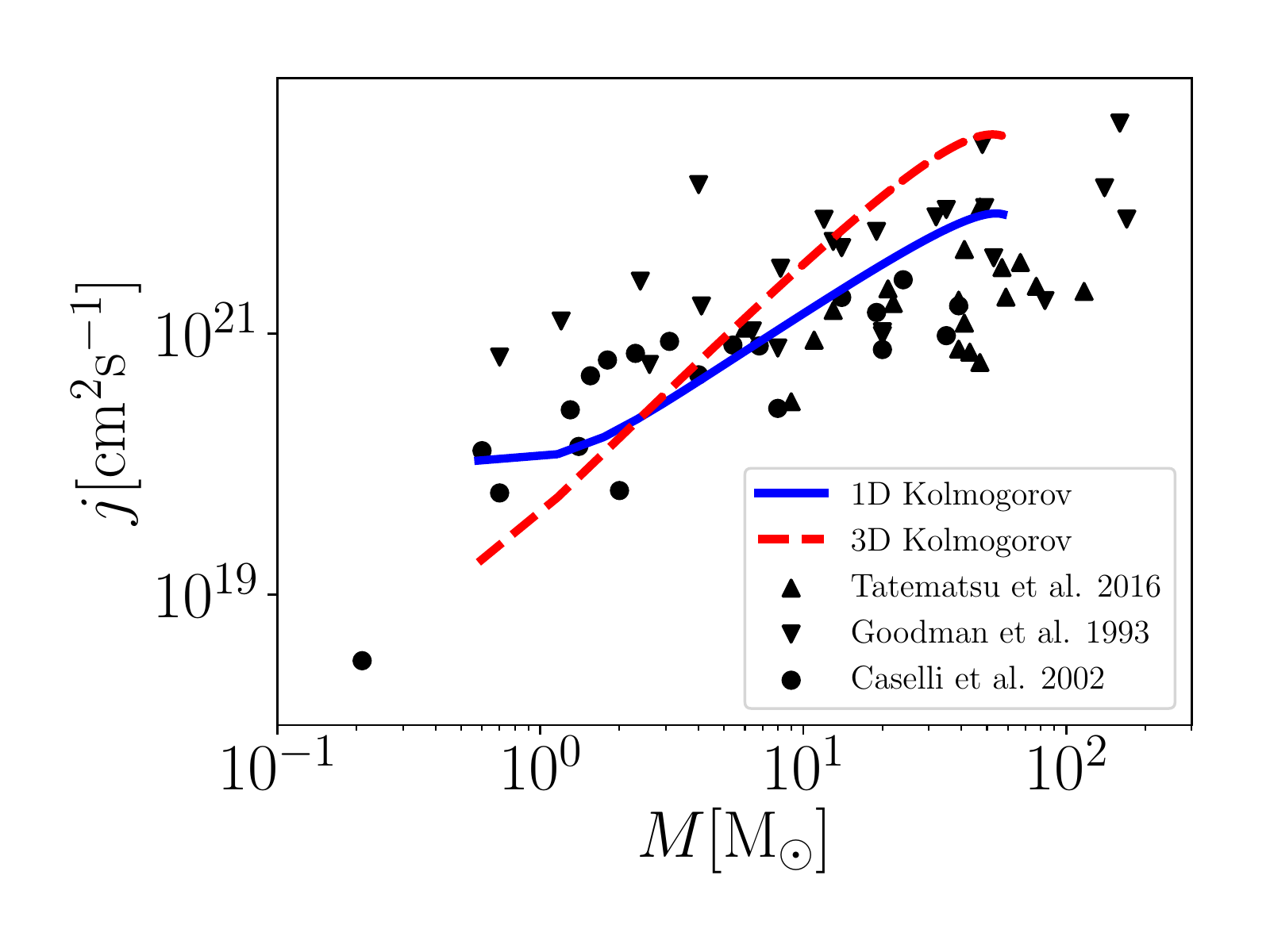}
\caption{$j-M$ relation for 1D Kolmogorov power spectrum. The red dashed line and blue solid line are 3D Kolmogorov power spectrum (same as Figure \ref{CAcfullkolonly}) and 1D Kolmogorov power spectrum model derived using \Eref{eq:1dkol}, respectively. The observational data are the same as in Figure \ref{goodmancasellitatematsu}.}
\label{AMjM5over3}
\end{figure}


\section{Comparison of the Derived Power Spectrum Models}
In this section, we attempt to evaluate the appropriateness of the different models presented above. To do so, we calculate $\sigma_{\rm error}^2$, the dispersion of the observational data points relative to the analytical $j-M$ relations for the different models. The dispersion of a model is defined as follows,
\begin{eqnarray}
\sigma^2_{\rm error} = \frac{1}{N_{\rm data}}\sum_{i=1}^{N_{\rm data}} (\log (j(M_i)) - \log(j_{{\rm obs},i}))^2,
\label{eq:dispdef}
\end{eqnarray}
where $i$, $N_{\rm data}$, $j(M_i)$, and $j_{{\rm obs},i}$ are each data point, the total number of data points, the analytical solution described by \Eref{ceq10}, and the observational points, respectively. Table \ref{table} shows the accuracy of the models normalized by that of the 3D Kolmogorov model. We give values of $\sigma^2_{\rm error}$ for the different models calculated with a velocity dispersion $\sigma_{\rm 3D}$ equal to $c_{\rm s}$ and $2c_{\rm s}$ corresponding to the observed range (cf. Section 3.1). Table \ref{table} shows that our conclusions are not affected by the choice of $\sigma_{\rm 3D}$ within the observed range. Note that the evaluation of appropriateness is tentative and future observations may give us more information to better constrain the models.

\begin{table}
\caption{\label{label}Relative appropriateness of the models. The values in Table \ref{table} are the dispersions $\sigma^2_{\rm error}$ of the observational points relative to the analytical solution in $j-M$ diagram calculated using \Eref{eq:dispdef} and normalized by $\sigma^2_{\rm error}$ of the 3D Kolmogorov model (cf., Section 4).} 
\begin{center}
\begin{tabular}{c|ccccc}
\br
 & 3D &  1D &Anisotropy & Log-normal &Compressible \\
 & Kolmogorov & Kolmogorov &  & ($\sigma_{\rm G}=0.65$, $\lambda_{\rm peak}=0.1$) & \\
\mr
$\sigma_{\rm 3D}=c_{\rm s}$ & 1 & 0.40 & 0.39 & 0.43 & 1.32\\ \mr
$\sigma_{\rm 3D}=2c_{\rm s}$ & 1 & 0.39 & 0.48 & 0.57 & 0.53\\
\br
\end{tabular}
\end{center}
\label{table}
\end{table}


\section{Discussion}
In the following, we attempt to provide a link between the velocity fluctuations along a star-forming filament and that of its surrounding molecular cloud. In particular we suggest that the 1D Kolmogorov velocity power spectrum along the filament described in Section 3.4 is linked to the 3D Kolmogorov velocity power spectrum of the parent cloud.

Recent observational results show that a small fraction of the total mass of molecular clouds ($< 20\%$) is in the form of dense gas, for column densities $\gtrsim 7 \times 10^{21}$ cm$^{-2}$, and that a large fraction of this dense gas is in the form of critical/supercritical filaments \citep[$\sim80\%$ on average, ][see also Torii et al. 2018]{Andre2014,Arzoumanian2019}. This column density value has been observationally identified as the threshold above which the star formation activity is significantly enhanced \citep[e.g.,][]{Heiderman2010, Lada2010, Lada2012, Konyves2015}. This star formation threshold is now understood as being equivalent to the line mass of 0.1 pc-wide filaments that is of the order of the critical value (${\rm M}_{\rm line,crit}$) for gravitational fragmentation \citep[e.g.,][]{Arzoumanian2013,Andre2014}. 

If we define the major axis of the filament as the $z$-axis ($x=y=0$), we can calculate the line mass of the filament as follows,
\begin{eqnarray}
M_{\rm line}(z) = \int \int^{\sqrt{x^2+y^2}<R_{\rm fil}} \rho \ dx dy.
\label{eq:disculinem1}
\end{eqnarray}
Likewise the velocity ${\bi v}_{\rm fil}$ along the filament is defined as follows,
\begin{eqnarray}
{\bi v}_{\rm fil}(z) = \frac{1}{M_{\rm line}}\int \int^{\sqrt{x^2+y^2}<R_{\rm fil}} \rho {\bi v} \  dx dy.
\label{eq:disculinem2}
\end{eqnarray}

The 1D Fourier component of the velocity field along the filament $z$-axis is 
\begin{eqnarray}
\widetilde{{\bi v}}(k_z) &= \frac{1}{L_z} \int^{L_z/2}_{-L_z/2} {\bi v}_{\rm fil}(z) \exp(- \rmi k_z z) dz \nonumber \\
&\sim \frac{1}{L_z} \int^{L_z/2}_{-L_z/2} {\bi v}(0,0,z) \exp(- \rmi k_z z) dz,
\label{eq:1dfourier}
\end{eqnarray}
where, $L_z$ is the length of the $z$ direction of the molecular cloud and $\widetilde{{\bi v}}$ is the 1D Fourier component of the velocity field. In the second step of \Eref{eq:1dfourier}, we take the limit $R_{\rm fil}\rightarrow 0$, because the mass of a critical/supercritical filament (i.e., the mass of dense gas) is much smaller than the total mass of the parental molecular cloud as suggested by observations (see above). The limit of ${\bi v}_{\rm fil}(z)$ as $R_{\rm fil}$ approaches to 0 is ${\bi v}(0,0,z)$. Note that while the velocity field is integrated with $x$ and $y$ in \Eref{eq:3DKPSalong}, in \Eref{eq:1dfourier} the velocity field is not integrated because of the approximation ${\bi v}_{\rm fil}(z) \sim {\bi v}(0,0,z)$. The 1D power spectrum $P_{\widetilde{{\bi v}}}$ of $\widetilde{{\bi v}}$ is defined as
\begin{eqnarray}
P_{\widetilde{{\bi v}}}(k_z) &= \left< | \widetilde{{\bi v}}(k_z) |^2 \right>.
\label{eq:1dfourier3}
\end{eqnarray}
Using \Eref{eq:velofieldfourier}, \Eref{eq:1dfourier}, and \Eref{eq:1dfourier3}, the 1D velocity power spectrum can be rewritten as 
\begin{eqnarray}
P_{\widetilde{{\bi v}}}(k_z) = \sum_{k_x'}\sum_{k_y'} P(k_x',k_y',k_z).
\label{eq:1dfourier4}
\end{eqnarray}
The derivation of \Eref{eq:1dfourier4} is given in Appendix B. The left hand side of \Eref{eq:1dfourier4} is the 1D power spectrum $P_{\widetilde{\bi v}}(k_z)$ of the velocity along the filament, which is equal to the integration of the 3D Kolmogorov-like power spectrum $P(k_x',k_y',k_z)$ with respect to $k_x'$ and $k_y'$. This integration can be done as follows,
\begin{eqnarray}
P_{\widetilde{{\bi v}}}(k_z) &= \sum_{k_x'}\sum_{k_y'} P(k_x',k_y',k_z) \nonumber \\
& \sim \frac{L_x L_y}{ (2\pi)^2} A \int^{k_{\rm max}}_{0} \int^{k_{\rm max}}_{0} (k_x'^2 + k_y'^2 + k_z^2)^{-11/6}\  dk_x' dk_y' \nonumber \\
& \sim \frac{L_x L_y}{ (2\pi)^2} A \int^{\infty}_{0} \int^{\infty}_{0} (k_x'^2 + k_y'^2 + k_z^2)^{-11/6}\  dk_x' dk_y' \nonumber \\
&= \frac{L_x L_y}{ (2\pi)^2} A \int^{\infty}_0 (k_r'^2 + k_z^2)^{-11/6} \ \frac{\pi}{2} k_r'dk_r' \nonumber \\
& \propto k_z^{-5/3},
\label{eq:1dfourier5}
\end{eqnarray}
where $L_x$ and $L_y$ are the length of $x$ and $y$ directions of the molecular cloud. $k_{\rm max}$ is the maximum wavenumber, which might correspond to an energy dissipation scale. The approximation that the summation can be replaced by an integral up to $+\infty$ is valid only for $k_z < k_{\rm max}$. Here we assume that the largest wavenumber is the same along all the three ($x$, $y$, and $z$) directions, i.e., isotropic at small scales. Therefore, we can use the approximation that the summation can be replaced by an integral up to $+\infty$. This result shows that the 1D approximation in \Eref{eq:1dfourier} leads to the 1D Kolmogorov power spectrum along the filament. While the width of the filament considered in Section 3.1 is also small (0.1 pc) compared to the size of the cloud ($\sim 10$ pc), the resulting slope of the power spectrum of ${\bi v}_{\rm 1D}(z)$ is $-11/3$, very different from the slope $-5/3$ found here. This is mainly due to the periodic boundary condition for the functional dependence of the modes in the $x$- and $y$-directions in Section 3.1: i.e., the modes in the $x$- and $y$-directions cancels in \Eref{eq:3DKPSalong}, these modes are not contributing to the power spectrum of ${\bi v}_{\rm 1D}(k_z)$, unlike in \Eref{eq:1dfourier4}. In this section, since we consider the velocity field that is extended in the parent cloud outside the filament, here the modes are not anymore periodic within the filament.

We can thus link the observed 3D power spectrum of the parent cloud to the 1D velocity power spectrum along the filament. A filament with ${\rm M}_{\rm line} \sim {\rm M}_{\rm line,crit}$ and velocity fluctuations along its crest characterized by the 1D Kolmogorov power spectrum with a slope of $-5/3$ would fragment into a series of cores presenting the observed distribution of angular momentum as shown in Section 3.4. 

In addition to providing an origin to the observed angular momentum of cores, we also emphasize that these velocity fluctuations along star-forming filaments with a power spectrum of $P(k) \propto k^{-5/3}$ may also be key in understanding the origin of the shape of the core mass function \citep[cf.,][]{Inutsuka2001,Roy2015,Lee2017}. Note also that the 1D approximation of star-forming filaments might be justified by the observed small mass fraction of the dense gas in the form of critical/supercritical filaments with respect to the total mass of the cloud. We suggest that this small mass fraction of dense gas may also provide a hint to the origin of the observed low star formation efficiency in molecular clouds.


\section{Summary}

In this paper we provide a causal link between properties of filaments and cores. We suggest that the observed angular momentum distribution of cores as a function of core mass can be understood by the fragmentation of filaments having velocity fluctuations close to the sonic speed and anisotropic, log-normal, and 1D Kolmogorov power spectra. We demonstrate that the 3D Kolmogorov velocity and density power spectra observed at the scale of the cloud implies a 1D Kolmogorov velocity spectrum along a filament when the mass of this latter is a small fraction of the total mass in the cloud, consistent with recent observational results \citep{Andre2014,Arzoumanian2019,Torii2018}. The results presented in this paper reinforces the key role of filament fragmentation in our understanding of the star formation process and the observed properties of cores. Future systematic observations tracing the velocity structure of cores are needed to derive the kinematic properties of cores to better constrain our models and examine the relation between the properties of filaments and cores, to understand the origin of the angular momentum of molecular cloud cores, multiple systems, and protoplanetary disks. 


\ack
We thank the referee for a constructive report which helped improving the clarity of the paper. Y. Misugi thanks T. Inoue and H. Kobayashi for their advice and encouragement. This work was supported by JSPS KAKENHI Grant Numbers 16H02160 \& 16F16024. D. Arzoumanian acknowledges an International Research Fellowship from the Japan Society for the Promotion of Science (JSPS). 


\appendix


\section{Derivation of $\bi{R}(\bi{k};L_{\rm core})$}
In this appendix, we show the detailed derivation of the Fourier component of the position $\bi{R}(\bi{k};L_{\rm core})$ of \Eref{Rdef}. Using the Bessel Functions of the first kind $J_B^{(1)}$, $J_B^{(2)}$ and the spherical Bessel functions $j_B^{(0)}$, $j_B^{(1)}$, the $x$-component of $\bi{R}(\bi{k};L_{\rm core})$ is
\begin{eqnarray}
\label{ceq5}
R_x(\bi{k};L_{\rm core}) =  - R_{\rm fil} \frac{k_x}{k_r}j_B^{(0)}(X)  \frac{J_B^{(2)}(Y)}{Y},
\end{eqnarray}
the $y$-component of $\bi{R}(\bi{k};L_{\rm core})$ is
\begin{eqnarray}
\label{ceq7}
R_y(\bi{k};L_{\rm core}) = - R_{\rm fil} \frac{k_y}{k_r}j_B^{(0)}(X)  \frac{J_B^{(2)}(Y)}{Y},
\end{eqnarray}
the $z$-component of $\bi{R}(\bi{k};L_{\rm core})$ is
\begin{eqnarray}
\label{ceq8}
R_z(\bi{k};L_{\rm core}) = - \frac{L_{\rm core}}{2} j_B^{(1)}(X) \frac{J_B^{(1)}(Y)}{Y},
\end{eqnarray}
where $X \equiv k_z L_{\rm core}/2$, $Y \equiv k_rR_{\rm fil}$, and $k_r = \sqrt{k_x^2 + k_y^2}$. The $k_x$, $k_y$, and $k_z$ are the wavenumber of $x$, $y$, and $z$ directions, respectively. To derive these expressions, we have used the following equations \citep{Abramowitz1965}:
\begin{eqnarray}
\label{ceq6}
\frac{d}{dx}[x^{-\nu}J_B^{({\nu})}(x)] = -x^{-\nu} J_B^{({\nu+1})}(x),
\end{eqnarray}

\begin{eqnarray}
\label{ceq9}
\frac{d}{dx}[x^{-\nu}j_B^{({\nu})}(x)] = -x^{-\nu} j_B^{({\nu+1})}(x),
\end{eqnarray}

\begin{eqnarray}
\label{ceq2}
j_B^{(0)}(X)=\frac{1}{L_{\rm core}}\int^{L_{\rm core}/2}_{-L_{\rm core}/2} dz \exp(\rmi k_z z),
\end{eqnarray}

\begin{eqnarray}
\label{ceq3}
J_B^{(0)}(Y)=\frac{1}{2 \pi}\int^{\pi}_{-\pi} d \phi  \exp(\rmi k_r r \cos \phi),
\end{eqnarray}

\begin{eqnarray}
\label{ceq4}
J_B^{(1)}(Y)=\frac{k_r}{R_{\rm fil}}\int_{0}^{R_{\rm fil}} J_B^{(0)}(k_rr)rdr,
\end{eqnarray}
where $\nu$ denotes the index of Bessel Function. In this paper we consider only integer values of $\nu$.

\section{Derivation of \Eref{eq:1dfourier3}}
In the following we detail the derivation of \Eref{eq:1dfourier4}. Using \Eref{eq:velofieldfourier}, \Eref{eq:1dfourier} can be written as,
\begin{eqnarray}
\widetilde{\bi v}(k_z) &= \frac{1}{L_z} \int^{L_z/2}_{-L_z/2} \sum_{k_x'}\sum_{k_y'}\sum_{k_z'} {\bi V}(\bi{k}') \exp(\rmi \bi{k}'\cdot \bi{x}) \exp(- \rmi k_z z) dz \nonumber \\
&= \frac{1}{L_z} \sum_{k_x'}\sum_{k_y'}\sum_{k_z'} {\bi V}(\bi{k}') \exp(\rmi k_x'x + \rmi k_y'y) \nonumber \\  &\quad \quad \times \int^{L_z/2}_{-L_z/2} \exp( \rmi k_z' z - \rmi k_z z) dz,
\label{eq:1dfourier2}
\end{eqnarray}
where $\widetilde{\bi v}(k_z)$ is the 1D Fourier component of the velocity field and ${\bi V}$ is the 3D Fourier component of the velocity field. Here, $x$ and $y$ are the cloud coordinates in the $x$-$y$ plane perpendicular to the filament axis. Since we take the limit $R_{\rm fil}\rightarrow 0$, we do not take the integration in the $x$-$y$ plane in this appendix (cf., \Eref{eq:disculinem2}). Since $\int^{L_z/2}_{-L_z/2} \exp( \rmi k_z' z - \rmi k_z z) dz = L_z \delta_{k_z,k_z'}$, where $\delta_{k_z,k_z'}$ is the Kronecker delta, \Eref{eq:1dfourier2} can be rewritten as
\begin{eqnarray}
\widetilde{\bi v}(k_z) =  \sum_{k_x'}\sum_{k_y'}\sum_{k_z'} {\bi V}(\bi{k}') \exp(\rmi k_x'x + \rmi k_y'y)  \delta_{k_z,k_z'}.
\label{eq:1dfourierapp0}
\end{eqnarray}
Finally, $\widetilde{\bi v}$ can be written as,
\begin{eqnarray}
\widetilde{\bi v}(k_z) = \sum_{k_x'}\sum_{k_y'} {\bi V}(k_x',k_y',k_z) \exp(\rmi k_x'x + \rmi k_y'y).
\label{eq:1dfourierapp2}
\end{eqnarray}
Using \Eref{powerspedefinewq} and \Eref{eq:1dfourierapp2}, we can derive \Eref{eq:1dfourier4}.

\bibliography{Angularmomentum_paper1_revise2}
\bibliographystyle{apj}

\end{document}